\documentclass[english,pra, a4paper, twocolumn, 10pt, showpacs]{revtex4}
\usepackage[T1]{fontenc}
\usepackage[latin9]{inputenc}
\usepackage{textcomp}
\usepackage{amsmath}
\usepackage{amssymb}
\usepackage{graphicx}

\makeatletter
\@ifundefined{textcolor}{}
{%
 \definecolor{BLACK}{gray}{0}
 \definecolor{WHITE}{gray}{1}
 \definecolor{RED}{rgb}{1,0,0}
 \definecolor{GREEN}{rgb}{0,1,0}
 \definecolor{BLUE}{rgb}{0,0,1}
 \definecolor{CYAN}{cmyk}{1,0,0,0}
 \definecolor{MAGENTA}{cmyk}{0,1,0,0}
 \definecolor{YELLOW}{cmyk}{0,0,1,0}
}

\makeatother

\usepackage{babel}
\begin{document}

\title{Improved numerical methods for infinite spin chains with long-range
interactions}

\author{V. Nebendahl}

\affiliation{Institut f\"ur Theoretische Physik, Universit\"at Innsbruck, Technikerstr.
25, A-6020 Innsbruck, Austria}

\author{W.D\"ur}

\affiliation{Institut f\"ur Theoretische Physik, Universit\"at Innsbruck, Technikerstr.
25, A-6020 Innsbruck, Austria}

\pacs{02.70.-c, 03.67.-a, 67.85.-d}
\begin{abstract}
We present several improvements of the infinite matrix product state
(iMPS) algorithm for finding ground states of one-dimensional quantum
systems with long-range interactions. As a main new ingredient we
introduce the superposed multi-optimization (SMO) method, which allows
an efficient optimization of exponentially many MPS of different length
at different sites all in one step. Hereby the algorithm becomes protected
against position dependent effects as caused by spontaneously broken
translational invariance. So far, these have been a major obstacle
to convergence for the iMPS algorithm if no prior knowledge of the
systems translational symmetry was accessible.
Further, we investigate some more general methods to speed up
calculations and improve convergence, which might be partially interesting in a much broader context, too.
As a more special problem, we also look into translational
invariant states close to an invariance braking phase transition
and show how to avoid convergence into wrong local minima for such systems.
Finally, we apply the new methods to polar bosons with long-range interactions.
We calculate several detailed Devil's Staircases with the corresponding phase diagrams
and investigate some supersolid properties.
\end{abstract}
\maketitle

\section{Introduction}

A numerical method for the simulation of large quantum systems needs
to meet two requirements: i)\nobreak \ an ansatz suitable for the
problem in question, and ii) efficient algorithms to find the (at least
nearly) optimal solution within the chosen ansatz. For one-dimensional
quantum systems on lattices, the currently most powerful numerical
tools are matrix product states (MPS) based algorithms including the
density matrix renormalization group (DMRG) \cite{Schollwoeck2011,White1992,Fannes1992,Affleck1987,MPSgleichDMRG}.
Their primary limitation is given by the amount of entanglement they
can handle. Several extensions of MPS have been conceived to overcome
this restriction as, (e.g.\  \cite{Vidal2007_MERA,Huebener2009RAGE,CTS}),
but for most practical applications MPS are still the first choice.
This is mainly due to the performance of the underlying optimization
routines. Although the general task to find ground states is known
to be NP hard \cite{Eisert2006NPhardDMRG,NPhardMPS}, commonly used
algorithms seem to have no problem to attain optimal MPS solutions
within computer precision for a plenitude of physical relevant systems.
Nevertheless, for some physical systems of interest these algorithms
still face severe difficulties.

In this paper, we treat such problematic cases given by ground states
of infinite spin chains with long-range interactions. An increasing
interest in reliable numerical methods for these states is, e.g.,\
triggered by the excellent experimental control of ultracold gases
and the possibility to realize systems with long-range dipole-dipole
interactions such as Rydberg atoms or polar molecules \cite{Burnell2009,Schachenmayer2010,UltraColdLongRange}.
Although these systems are of finite size, one is often interested
in the thermodynamical limit (i.e.,\ infinite systems) for a better
insight.

Different strategies are known for the numerical study of ground states
in the thermodynamic limit. One might try to extrapolate results from
a series of increasingly large finite systems \cite{FiniteSizeScaling,Pirvu_Scaling,FiniteSizeEffektSuppressing}
or directly construct the infinite state itself. The latter is, e.g.,\ done
by the infinite time-evolving block decimation (iTEBD) algorithm \cite{TEBD},
which is based on an explicit translational invariant ansatz. This
ansatz is quite elegant for interactions which are restricted to nearest
neighbors, while it gets impractical for long-range interactions.

A comfortable way to incorporate long-range interactions is to encode
them into a matrix product operator (MPO) \cite{McCulloch2007_HamMPO,PolyExp,Crosswhite2008_Automata,Froewis2010}.
This concept can be integrated in an infinite matrix product state
(iMPS) algorithm \cite{Crosswhite2008,McCulloch}.

The basic idea of this iMPS algorithm is to obtain the ground state
of an infinite system as the fixed point of a constantly growing finite
state by inserting iteratively new sites into its middle until convergence
is reached. A major disadvantage of this approach is that the algorithm
generally fails to converge if the ground state has a non-trivial
translational symmetry.

In this paper, various extensions to the basic iMPS algorithm are presented. As a central element, the superposed multi-optimization (SMO) is introduced (section~\ref{sub:Parallel-optimization}), which provides a remedy for the just mentioned convergence problem. Here, the key idea is to join the optimization of exponentially many MPS in a superposition and solve it efficiently. Due to this superposed optimization, the effective overall problem becomes translational invariant again and poses no longer a hindrance for convergence.

We also introduce several improvements which work independently of the SMO method. As such, we present two different modifications of the MPS optimization routine: On the one hand, we use  physical insight for systems close to translational invariance breaking phase transition to suggest  a method to reduce the danger of  being trapped into a local minimum of the energy (section~\ref{sub:Translational-invariant-ground-1}). On the other hand, we provide a more technical discussion how to recycle information from previous optimizations  to speed up calculations (section~\ref{sub:Minimization-routine-and}). As a part of these considerations we provide a simple implementation of a Davidson like algorithm \cite{DavidsonJacobi} based on information from previous optimizations (appendix~\ref{sub:Appendix:-Davidson-implementation}). This implementation is not bonded to the MPS framework and hence might be used in much broader context.

All algorithms are described in depth, such that readers who are willing to reproduce our results should find all the information needed for successful programming.
As a consequence, readers who just like to understand  the crucial ideas might find the amount of algorithmic details far beyond their interest.
Having these two types of readers in mind, we partitioned the material according to its level of detail into different chapters, such that entire passages can be omitted.

In section~\ref{sec:Basic-algorithm}, we review the basic concepts and the iMPS algorithm as presented in \cite{Crosswhite2008}. Readers who feel save to skip this part find in Fig.~\ref{fig:MPS} and Fig.~\ref{fig:MPO} a pictorial description of all symbols used in this section.
In Section~\ref{sec:broken-translational-invariance}, the main new concepts of this paper are presented, above all the SMO method. Changes of the algorithm are kept to a minimum in here in contrary to section~\ref{sec:Enhanced-algorithm} where several algorithmic improvements and their numerical realization are presented in detail.
Each subsections of section~\ref{sec:Enhanced-algorithm} contains an individual topic, which is presented in the first lines. These subsections can be skipped without danger of losing the ability to understand the rest of the paper. A slight exception might be section~\ref{sub:Minimization-routine-and}, in which the concept of iterative eigenvector solver based on subspace projections is reviewed. Familiarity with this concept is assumed in section~\ref{sub:Altered-minimization-routine}, \ref{sub:LengthOfMPS} and appendix~\ref{sub:Subspace-projection-and}.
In section \ref{sec:Applications}, we apply our new algorithm to a system of polar bosons with long-range interactions. Detailed calculations of Devil's Staircases and phase diagrams are shown and a supersolid like phase is investigated. Finally, the paper is complemented by an appendix, into which several details have been outsourced.

\section{Basic algorithm\label{sec:Basic-algorithm}}

In this section we review fundamental concepts \cite{Schollwoeck2011}
and the iMPS algorithm as presented in reference \cite{Crosswhite2008}.
For this algorithm to work not only the Hamiltonian but also the ground
state have to be translational invariant. The extension to ground
states with broken translational symmetry will be introduced in section
\ref{sec:broken-translational-invariance}.

\subsection{MPS and MPO\label{sub:MPO}}

In this paper we deal with spin chains. The quantum state of a spin
chain is determined by the inner degrees of freedom of its components

\begin{equation}
|\psi\rangle=\sum_{s_{1}s_{2}\cdots s_{n}}\mathcal{A}_{s_{1}s_{2}\cdots s_{n}}\cdot|s_{1}\rangle\otimes|s_{2}\rangle\otimes\cdots\otimes|s_{n}\rangle.\label{eq:Psi}
\end{equation}
Since the size of the tensor $\mathcal{A}_{s_{1}s_{2}\cdots s_{n}}$
grows exponentially with the number of sites, a more economical representation
is needed. A matrix product state (MPS) \cite{Schollwoeck2011,Fannes1992,Affleck1987}
consists in the ansatz

\begin{eqnarray}
\mathcal{A}_{s_{1}s_{2}\cdots s_{n}} & = & A_{[1]\, s_{1}}^{\alpha_{1}}\cdot A_{[2]\, s_{2}}^{\alpha_{1}\alpha_{2}}\cdot A_{[3]\, s_{3}}^{\alpha_{2}\alpha_{3}}\dots\nonumber \\
 &  & \cdots A_{[n-1]\, s_{n-1}}^{\alpha_{n-2}\alpha_{n-1}}\cdot A_{[n]\, s_{n}}^{\alpha_{n-1}},\label{eq:MPS}
\end{eqnarray}
where we used the Einstein summation convention. For a general exact
quantum state exponentially growing bond dimensions $\alpha_{i}$
are needed, but even for infinite systems excellent approximations
are possible with moderate bond dimensions if the ground state fulfills
an area law for the entanglement entropy \cite{Hastings2007}.

For Hamiltonians
\begin{equation}
\hat{H}=\sum_{s_{1}\cdots s_{n},s'_{1}\cdots s'_{n}}\mathcal{H}_{s_{1}\cdots s_{n}}^{s'_{1}\cdots s'_{n}}\cdot|s'_{1}\rangle\langle s_{1}|\otimes\cdots\otimes|s'_{n}\rangle\langle s_{n}|\label{eq:HamiltonBasis}
\end{equation}
a similar ansatz as for the quantum state (\ref{eq:MPS}) leads to
the concept of matrix product operators (MPO) \cite{McCulloch2007_HamMPO,PolyExp,Crosswhite2008_Automata,Froewis2010}

\begin{eqnarray}
\mathcal{H}_{s_{1}s_{2}\cdots s_{n}}^{s'_{1}s'_{2}\cdots s'_{n}} & = & H_{[1]\, s'_{1}s_{1}}^{\mu_{1}}\cdot H_{[2]\, s'_{2}s_{2}}^{\mu_{1}\mu_{2}}\cdot H_{[3]\, s'_{3}s_{3}}^{\mu_{2}\mu_{3}}\dots\nonumber \\
 &  & \cdots H_{[n-1]\, s'_{n-1}s_{n-1}}^{\mu_{n-2}\mu_{n-1}}\cdot H_{[n]\, s'_{n}s_{n}}^{\mu_{n-1}}.\label{eq:MPO}
\end{eqnarray}
Many relevant Hamiltonians are represented by MPO with relative small
bond dimensions. For our purposes, it is important to mention that
this is often also true for Hamiltonians with long-range interaction
terms (see, e.g.,\ \cite{Froewis2010}). A recipe for the explicit
construction is explained in Appendix \emph{\ref{sub:Appendix:-MPO-representation}.}
In the case of a translational invariant Hamiltonians (see remark
below), the MPO can be built in such a fashion that all tensors \emph{$H_{[i]\, s'_{i}s_{i}}^{\mu_{i-1}\mu_{i}}$
}are identical for \emph{$2\leq i\leq n-1$.} This allows us to drop
the index in the square brackets except for the leftmost and rightmost
tensors.

\begin{eqnarray}
\mathcal{H}_{s_{1}s_{2}\cdots s_{n}}^{s'_{1}s'_{2}\cdots s'_{n}} & = & H_{[L]\, s'_{1}s_{1}}^{\mu_{1}}\cdot H_{s'_{2}s_{2}}^{\mu_{1}\mu_{2}}\cdot H_{s'_{3}s_{3}}^{\mu_{2}\mu_{3}}\cdot H_{s'_{4}s_{4}}^{\mu_{3}\mu_{4}}\dots\nonumber \\
 &  & \cdots H_{s'_{n-2}s_{n-2}}^{\mu_{n-3}\mu_{n-2}}\cdot H_{s'_{n-1}s_{n-1}}^{\mu_{n-2}\mu_{n-1}}\cdot H_{[R]\, s'_{n}s_{n}}^{\mu_{n-1}}\label{eq:TransMPO}
\end{eqnarray}
Further more, all tensors $H$ are independent of the total number
of sites. As a consequence, the MPO of a translational invariant Hamiltonian
for $n$ sites can easily be augmented to $n+1$ sites by just inserting
another tensor $H$.

Apart from an efficient representation of quantum states and operators,
MPS and MPO also provide an efficient way to calculate expectation
values. For details, we refer to the literature (as, e.g.,\ \cite{Schollwoeck2011}).

\textbf{Remark:} In this paper we apply the term \emph{translational
invariant} also to finite systems (with open boundary conditions)
and their Hamiltonians which are used in the iMPS algorithm to approach
the infinite case.

\subsection{Overview of the iMPS algorithm\label{sub:The-idea-behind}}

Any algorithm which deals with infinite MPS could be addressed as
iMPS algorithm. In this paper we use this term exclusively for algorithms
of the type as presented in reference \cite{Crosswhite2008}. This
algorithm aims at finding an MPS representation for the ground state
of an infinite one-dimensional quantum system. In its plain version
the iMPS algorithm takes translational invariance for granted such
that all sites behave equally. Thus all we need to construct the entire
state is a perfect description of one site and its entanglement features
with its environment given by the rest of the system. This environment
is dominated by nearby neighbor sites while the influence of sites
far away can be neglected in any one-dimensional system with an asymptotic
decay of correlations faster than $r^{-1}$. Therefore the environment
built up by an infinite system can be simulated with a finite system.
Correspondingly the center site of a sufficiently large but finite
system provides a good approximation for its counterpart in the infinite
case.

The iMPS algorithm is built upon a finite system, which is iteratively
enlarged by inserting new sites into its middle. Since we express
quantum states by MPS each of these new sites is represented by an
individual tensor $A_{[n]}$ \eqref{eq:MPS}. Before a new tensor
$A_{[n]}$ is inserted it is optimized such that the resulting energy
\begin{equation}
E=\frac{\langle\psi|H|\psi\rangle}{\langle\psi|\psi\rangle}
\end{equation}
 is minimized. Hereby all previously inserted tensors $A_{[j<n]}$
are left unchanged. Of course, these local optimizations of the new
tensors $A_{[n]}$ are generally not sufficient to find the lowest
energy state of the \emph{entire }system. What we are supposed to
get is a ground state approximation which might be bad at the outer
edges but close to the center, it should become better with each new
tensor inserted. This is all we need to obtain an adequate description
of the center site's environment, since the influence of the outer
sites fades away with distance, anyway. Therefore, we expect the environment
of the center site to converge towards its infinite counterpart, and
with that the new tensors $A_{[n]}$ inserted into each round should converge,
too

\begin{equation}
A_{[n]}\rightarrow A_{\textrm{[converged]}}.\label{eq:Convergence Limes}
\end{equation}

For ground states which violate translational invariance, it is no
longer given that all sites behave equally. At this point, our intuitive
argumentation breaks down and the algorithm generally fails to converge.
We will study this case in section \ref{sec:broken-translational-invariance}.
For the time being, we assume that the algorithm ends up with a converged
tensor $A_{\textrm{[converged]}}$. With the help of this one tensor,
the entire iMPS can be constructed using the rules we will encounter
in \ref{sub:Constructing-the-MPS}.

\subsubsection{Long-range interactions}

An important ingredient for a fast computer code is a clever book-keeping
of the interaction terms, which allows us to save many calculations due
to recycling. In the case of long-range interactions, this task becomes
tricky, since the iMPS algorithm permanently splits the MPS and adds
new sites. By this means, the distances between sites on the left
and right halves change each time and with them all distance-dependent
interaction terms. Nonetheless, for translational invariant Hamiltonians
recycling can still be done in a well-arranged fashion by encoding
the Hamiltonian into an MPO as in equation \eqref{eq:TransMPO}. Every
time the MPS is enlarged by a new tensor, the MPO is enlarged by the
standard tensor $H_{s's}^{\mu_{l}\mu_{r}}$, too. This simple procedure
automatically corrects all distance-dependent interaction terms.

\subsection{Constructing the MPS\label{sub:Constructing-the-MPS}}

\begin{figure}
\includegraphics[width=1\columnwidth]{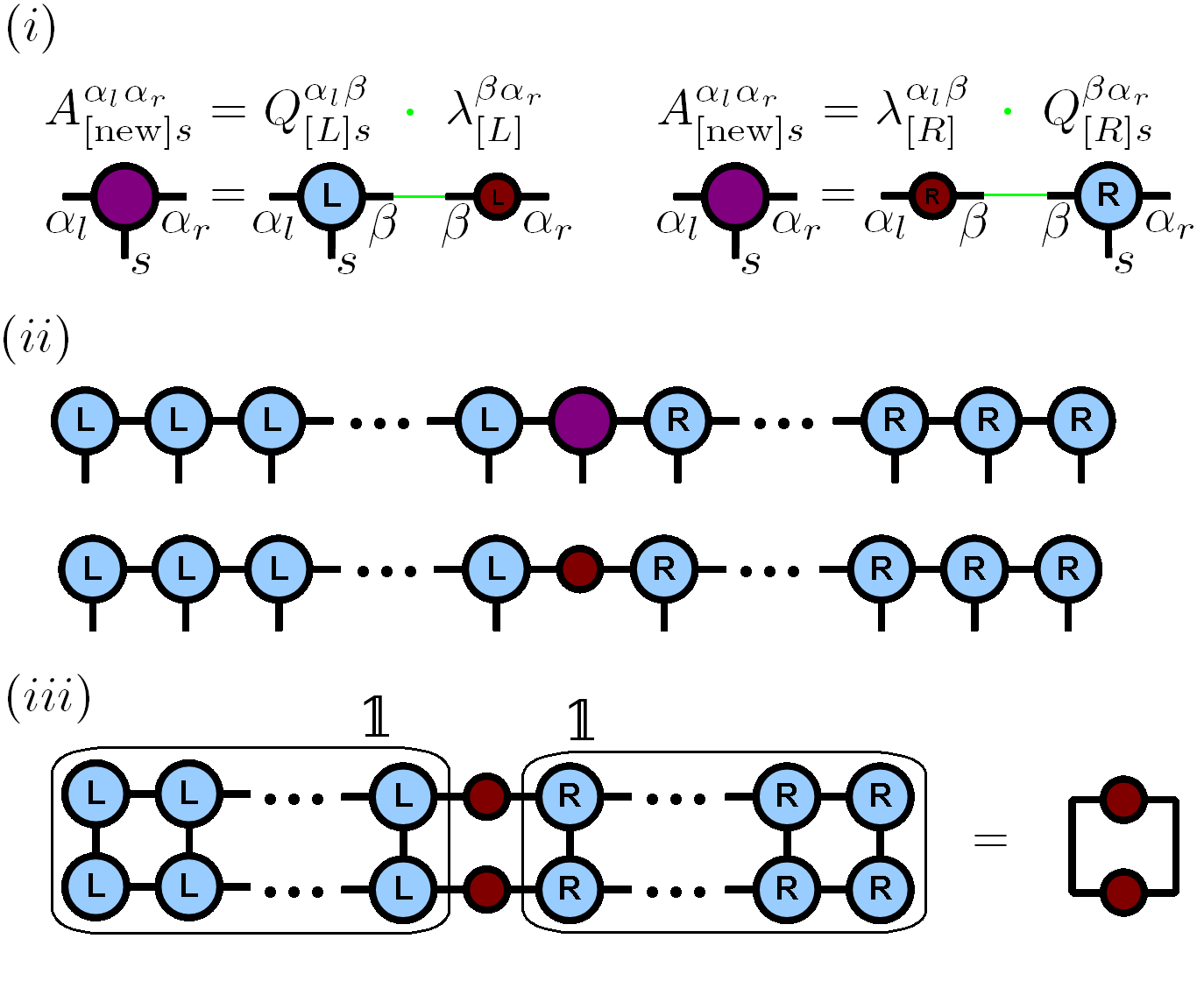}\caption{$(i)$ Diagrammatic representation of the two decompositions of $A$
according to equation \eqref{eq:Decomposition}. Vertical legs correspond
to physical indices while horizontal legs belong to the auxiliary
indices. Connected legs are summed over. $(ii)$ Resulting structure
of the MPS \eqref{eq:Q-MPS} before and after the decomposition of
the last $A_{\textrm{[new]}}$. $(iii)$ MPS version of $\langle\psi|\psi\rangle$,
where the turned over MPS symbolizes the complex conjugate. Due to
the orthogonal decomposition \eqref{eq:SVD} the tensors in the left
and right boxes generate the identity \eqref{eq:Orthogonal}, which allows
an easy control of the MPS norm~\eqref{eq:MPS-Norm}.\label{fig:MPS}}

\end{figure}

So far, we just mentioned that the iMPS algorithm constructs the MPS
by constantly inserting new tensors in its middle. We will now specify
on that. First, it is convenient to treat the MPS as divided into
two halfs, left and right from the center. Each time the optimization
procedure described in \ref{sub:Tensor-optimization} provides a new
tensor $A_{s}^{\alpha_{l}\alpha_{r}}$, we have to decide into which
half $A_{s}^{\alpha_{l}\alpha_{r}}$ is to be absorbed. This is done
under the following rules:
\begin{enumerate}
\item According to the half in which $A_{s}^{\alpha_{l}\alpha_{r}}$ is
to be absorbed, decompose it (Fig.\,\ref{fig:MPS}\,$(i)$) as
\begin{equation}
A_{s}^{\alpha_{l}\alpha_{r}}=\begin{cases}
Q_{[L]\, s}^{\alpha_{l}\beta}\cdot\lambda_{[L]}^{\beta\alpha_{r}} & \textrm{left half}\\
\lambda_{[R]}^{\alpha_{l}\beta}\cdot Q_{[R]\, s}^{\beta\alpha_{r}} & \textrm{right half}
\end{cases}.\label{eq:Decomposition}
\end{equation}
$Q$ is orthogonalized such that
\begin{equation}
Q_{[L]\, s}^{\alpha_{l}\beta}\cdot Q_{[L]\, s}^{*\alpha_{l}\beta'}=\delta^{\beta\beta'}=Q_{[R]\, s}^{\beta\alpha_{r}}\cdot Q_{[R]\, s}^{*\beta'\alpha_{r}},\label{eq:OrthogonalQ}
\end{equation}
where the asterisk denotes the complex conjugation (see the upcoming
equation (\ref{eq:SVD}) for details).
\item With each new tensor $A$ overwrite the $\lambda$ of the tensor before.
\end{enumerate}
Each $A_{[n]}$ is optimized in such a fashion that it compensates
for the overwritten $\lambda_{[n-1]}$.

For the orthogonalization \eqref{eq:OrthogonalQ} we proceed as follows:
In a first step, we write the tensor $A_{s}^{\alpha_{l}\alpha_{r}}$
as matrix $A^{m,a}$ where $m$ is a multi-index. If the tensor $A_{s}^{\alpha_{l}\alpha_{r}}$
is to be absorbed into the left MPS half $m=(s,\alpha_{l})$ elsewise
$m=(s,\alpha_{r})$. For the leftmost and rightmost tensors of the
MPS, which have only two indices, $m=s$. The index $a$ corresponds
to the leftover index of $A_{s}^{\alpha_{l}\alpha_{r}}$, which is
not in $m$. Next, the matrix $A^{m,a}$ is decomposed into an orthogonal
part $Q$ and a ``rest'' part $\lambda$. Different decompositions
would fulfill this task, but for the working of the algorithm it is
best to resort to a singular value decomposition
\begin{eqnarray}
A & = & U\cdot D\cdot V^{\dagger}\nonumber \\
 & = & \underbrace{U\cdot V^{\dagger}}_{Q}\cdot\underbrace{V\cdot D\cdot V^{\dagger}}_{\lambda}.\label{eq:SVD}
\end{eqnarray}
Rewritten as tensors we end up with equation (\ref{eq:Decomposition}).

A consequent application of these rules yields an MPS built of orthogonalized
tensors $Q$ and one single matrix $\lambda$ from the very last $A_{\textrm{[new]}}$
in the center (Fig.\,\ref{fig:MPS}\,$(ii)$).
\begin{equation}
Q_{[L]\, s_{1}}^{[1]\,\alpha_{1}}\cdots Q_{[L]\, s_{k}}^{[k]\,\alpha_{k-1}\widetilde{\alpha}_{k}}\cdot\lambda^{\widetilde{\alpha}_{k}\alpha_{k}}\cdot Q_{[R]\, s_{k+1}}^{[k+1]\,\alpha_{k}\alpha_{k+1}}\cdots Q_{[R]\, s_{n}}^{[n]\,\alpha_{n-1}}\label{eq:Q-MPS}
\end{equation}
This MPS standard form is numerical robust and has an easily calculated
norm $\Vert\psi\Vert=\sqrt{\langle\psi|\psi\rangle}$. To see this
let us multiply $Q_{[L]\, s_{1}}^{[1]\,\alpha_{1}}\cdots Q_{[L]\, s_{k}}^{[k]\,\alpha_{k-1}\alpha_{k}}$
(the left half of equation \eqref{eq:Q-MPS}) with its complex conjugated

\begin{align}
\left(Q_{[L]\, s_{1}}^{[1]\,\alpha_{1}}\cdot Q_{[L]\, s_{2}}^{[2]\,\alpha_{1}\alpha_{2}}\cdot\dots\right)\cdot\left(Q_{[L]\, s_{1}}^{*[1]\,\beta1}\cdot Q_{[L]\, s_{2}}^{*[2]\,\beta_{1}\beta_{2}}\cdot\dots\right) & =\nonumber \\
\underbrace{\underbrace{\underbrace{\left(Q_{[L]\, s_{1}}^{[1]\,\alpha_{1}}\cdot Q_{[L]\, s_{1}}^{*[1]\,\beta_{1}}\right)\cdot}_{\delta^{\alpha_{1}\beta_{1}}}\left(Q_{[L]\, s_{2}}^{[2]\,\alpha_{1}\alpha_{2}}\cdot Q_{[L]\, s_{2}}^{*[2]\,\beta_{1}\beta_{2}}\right)}_{\overset{\delta^{\alpha_{2}\beta_{2}}}{\vdots}}\dots}_{\delta^{\alpha_{k}\beta_{k}}}\label{eq:Orthogonal}
\end{align}
and analog for the right half. Thanks to equation \eqref{eq:OrthogonalQ}
all $Q\cdot Q^{*}$-pairs turn into $\delta$-functions and the norm
of the MPS (\ref{eq:Q-MPS}) equals the remaining $\Vert\lambda\Vert$
(see Fig.\,\ref{fig:MPS}\,$(iii)$) which also equals the norm
of the last inserted tensor $A_{\textrm{[new]}}$. Thus
\begin{equation}
\Vert\textrm{MPS}\Vert=\Vert A_{\textrm{[new]}}\Vert.\label{eq:MPS-Norm}
\end{equation}

\subsection{Tensor optimization\label{sub:Tensor-optimization}}

The iMPS algorithm is an iterative procedure. As described in section
\ref{sub:The-idea-behind} new tensors $A_{[n]}$ are constantly inserted
into the MPS, which represents the finite state $\psi$. Each of these
new tensors $A_{[n]}$ is optimized such that the energy $E=\frac{\langle\psi|H|\psi\rangle}{\langle\psi|\psi\rangle}$
is minimized. As we will discuss in more detail below, this can be
written as
\begin{equation}
\min\frac{\langle\psi|H|\psi\rangle}{\langle\psi|\psi\rangle}\rightarrow\min\frac{\langle A_{[n]}|\mathbb{H}_{[n]}|A_{[n]}\rangle}{\langle A_{[n]}|\mathbb{I}_{[n]}|A_{[n]}\rangle}.\label{eq:Effektive-Min}
\end{equation}

\begin{itemize}
\item $|A_{[n]}\rangle$ is the vectorized form of the tensor $A_{[n]}$,
that is $|A_{[n]}\rangle=A_{[n]}^{i}=A_{[n]\, s}^{\alpha_{l}\alpha_{r}}$
with the multi-index $i=(\alpha_{l},\alpha_{r},s).$
\item $\mathbb{H}_{[n]}$ is an effective operator built of the MPO representation
of the Hamiltonian $H$ \eqref{eq:TransMPO} and all MPS tensors of
$\langle\psi|$ and $|\psi\rangle$ except for the two new $A_{[n]}$.
\item $\mathbb{I}_{[n]}$ is the identity operation thanks to the orthogonalized
standard form of the MPS, see \eqref{eq:Orthogonal}.
\end{itemize}
Equation \eqref{eq:Effektive-Min} is solved by setting $|A_{[n]}\rangle$
equal to the lowest eigenvector of $\mathbb{H}_{[n]}$.

\textbf{Remark:} We will repeatedly use the notation $|T\rangle$
or $\langle T|$ for a vectorized tensor $T$.

\subsubsection{Effective operator $\mathbb{H}$}

\begin{figure}
\includegraphics[width=1\columnwidth]{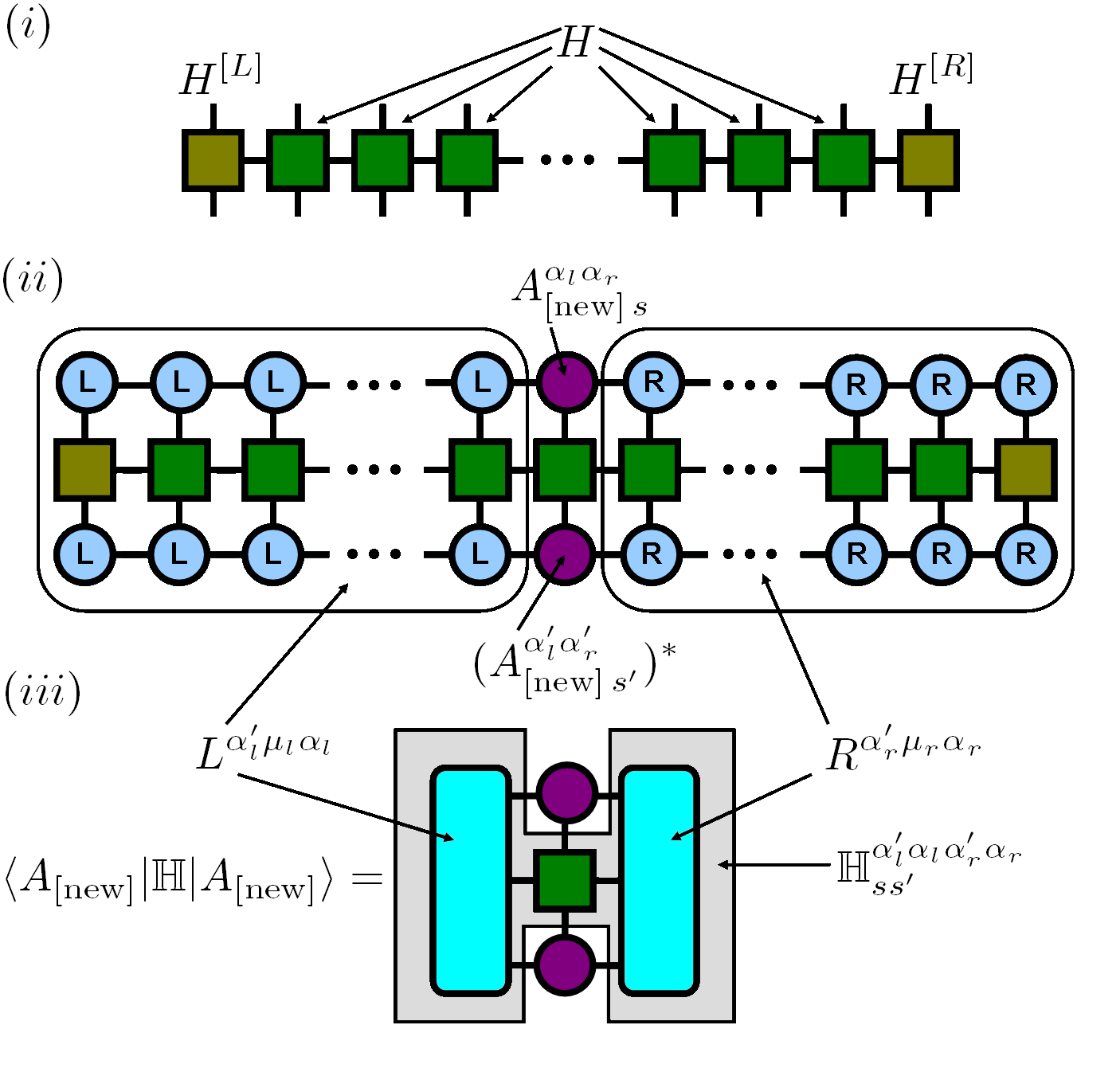}\caption{$(i)$ Diagrammatic representation of the Hamiltonian MPO \eqref{eq:TransMPO}.
$(ii)$ MPS and MPO realization of $\langle\psi|H|\psi\rangle$ (compare
with Fig.\,\ref{fig:MPS}). The boxes indicate the left and right
halfs $L^{\alpha'_{l}\mu_{l}\alpha_{l}}$ \eqref{eq:New Left-Half}
and $R^{\alpha'_{r}\mu_{r}\alpha_{r}}$ \eqref{eq:New Right-Half}.
$(iii)$ Same object as above with contracted inner indices of $L^{\alpha'_{l}\mu_{l}\alpha_{l}}$
and $R^{\alpha'_{r}\mu_{r}\alpha_{r}}$. The object in the H-shaped
box corresponds to the effective operator $\mathbb{H}$ \eqref{eq:Effektive H}.\label{fig:MPO}}
\end{figure}

In \ref{sub:Constructing-the-MPS} we mentioned that it is convenient
to treat the MPS as divided into two halves: left and right
from the newest tensor $A_{[n]}$. For the same reason we decompose
$\mathbb{H}_{[n]}$ into a left half $L_{[n]}^{\alpha'_{l}\mu_{l}\alpha_{l}}$
and a right half $R_{[n]}^{\alpha'_{r}\mu_{r}\alpha_{r}}$, which
are connected by a single MPO tensor $H_{s_{n}'s_{n}}^{\mu_{l}\mu_{r}}$
\eqref{eq:TransMPO} corresponding to the new site (see Fig.\,\ref{fig:MPO}\,$(iii)$).
\begin{equation}
\mathbb{H}_{[n]}^{i',i}=\mathbb{H}_{[n]\, s'_{n}s_{n}}^{\alpha'_{l}\alpha_{l}\alpha'_{r}\alpha_{r}}=L_{[n]}^{\alpha'_{l}\mu_{l}\alpha_{l}}\cdot H_{s'_{n}s_{n}}^{\mu_{l}\mu_{r}}\cdot R_{[n]}^{\alpha'_{r}\mu_{r}\alpha_{r}}\label{eq:Effektive H}
\end{equation}
with $i=(\alpha_{l},\alpha_{r},s_{n})$.

Since the iMPS algorithm is an iterative procedure, $L_{[n]}$ and
$R_{[n]}$ are built up iteratively, as well. Suppose we intend to
absorb the $A_{[n-1]}$ of the previous optimization step into the
left half. First, we use equation \eqref{eq:Decomposition} to gain
the orthogonal tensor $Q_{[L]\, s_{n-1}}^{\alpha_{\bar{l}}\alpha_{l}}$.
With that
\begin{equation}
L_{[n]}^{\alpha'_{l}\mu_{l}\alpha_{l}}=L_{[n-1]}^{\alpha'_{\bar{l}}\mu_{\bar{l}}\alpha_{\bar{l}}}\cdot Q_{[L]\, s'}^{*\:\alpha'_{\bar{l}}\alpha'_{l}}\cdot H_{s',s}^{\mu_{\bar{l}}\mu_{l}}\cdot Q_{[L]\, s}^{\alpha_{\bar{l}}\alpha_{l}},\label{eq:New Left-Half}
\end{equation}
where the asterisk denotes complex conjugation. In this case, where
the tensor $A_{[n-1]}$ is absorbed into the left half, the right
half stays unchanged $R_{[n]}=R_{[n-1]}$. Obverse, if we decide to
absorb $A_{[n-1]}$ into the right half,  the left half stays unchanged
and $R$ becomes
\begin{equation}
R_{[n]}^{\alpha'_{r}\mu_{r}\alpha_{r}}=Q_{[R]\,\alpha'_{r}\alpha'_{\bar{r}}}^{*\:\alpha'_{r}\alpha'_{\bar{r}}}\cdot H_{s',s}^{\mu_{r}\mu_{\bar{r}}}\cdot Q_{[R]\, s}^{\alpha_{r}\alpha_{\bar{r}}}\cdot R_{[n-1]}^{\alpha'_{\bar{r}}\mu_{\bar{r}}\alpha_{\bar{r}}}.\label{eq:New Right-Half}
\end{equation}

\subsection{Algorithm\label{sub:Algorithm}}

After having presented the decisive ingredients of the iMPS algorithm,
we like to emphasize the steps one actually has to perform on the
computer. The algorithm consists of an initializing procedure (see
\ref{sub:Initialization}) and an iteration loop, which is repeated
until convergence is reached $A_{[n]}\rightarrow A_{[\textrm{converged}]}$
\eqref{eq:Convergence Limes}. The tensor $A_{[\textrm{converged}]}$
is all we need to calculate expectation values. We do not hold any
copy of the MPS we are calculating. The only objects stored (in the
purest version of the algorithm) are the actual versions of $L_{[n]}^{\alpha'_{l}\mu_{l}\alpha_{l}}$
\eqref{eq:New Left-Half}, $R_{[n]}^{\alpha'_{r}\mu_{r}\alpha_{r}}$
\eqref{eq:New Right-Half} and $A_{[n]\, s}^{\alpha_{l}\alpha_{r}}$.

\subsubsection{Loop\label{sub:Loop}}
\begin{enumerate}
\item Calculate the new $A_{[n]\, s}^{\alpha_{l}\alpha_{r}}=A_{[n]}^{i}$
with $i=(\alpha_{l},\alpha_{r},s)$. Therefor

\begin{enumerate}
\item Use \eqref{eq:Effektive H} to calculate $\mathbb{H}_{[n]}^{i',i}=\mathbb{H}_{[n]\, s,s'}^{\alpha'_{l}\alpha_{l}\alpha'_{r}\alpha_{r}}$.
\item Set $A_{[n]}^{i}$ equal to the lowest eigenvector of $\mathbb{H}_{[n]}^{i',i}$
\end{enumerate}
\item Decide, whether to absorb $A_{[n]}^{i}$ into the left or right half
(e.g.\  even steps left, odd steps right) and act accordingly in
the following two steps.
\item Use \eqref{eq:Decomposition} and \eqref{eq:SVD} to decompose $A_{[n]}^{i}$
and gain $Q_{[L]}$ or $Q_{[R]}$.
\item Use \eqref{eq:New Left-Half} or \eqref{eq:New Right-Half} to get
$L_{[n+1]}^{\alpha'_{l}\mu_{l}\alpha_{l}}$ and $R_{[n+1]}^{\alpha'_{r}\mu_{r}\alpha_{r}}$.
\end{enumerate}

\subsubsection{Initialization\label{sub:Initialization}}

At the beginning we have to initialize the values of $L^{\alpha'_{l}\mu_{l}\alpha_{l}}$
and $R^{\alpha'_{r}\mu_{r}\alpha_{r}}$. This can be done with the
help of an exact solution $\psi$ for a small system best with an
even number of sites $n=2k$ (we mostly used $n=8\:\textrm{or}\:10$
for two-level sites). The state $\psi$ is split into a left and right
part, which allows to calculate $L$ and $R$. In more details, note the following:
\begin{enumerate}
\item Write the Hamiltonian of the small system as matrix with multi-indices
$H^{\left(s'_{1}\cdots s'_{n}\right),\left(s_{1}\cdots s_{n}\right)}$
and solve for $\psi$$^{\left(s_{1}\cdots s_{n}\right)}$.
\item Split the multi-index in two multi-indices, giving $\psi^{\left(s_{1}\cdots s_{k}\right),\left(s_{k+1}\cdots s_{n}\right)}$,
which can be interpreted as a matrix.
\item Use a singular value decomposition $\psi=U\cdot D\cdot V^{\dagger}$
(or Takagi's factorization $\psi=U\cdot D\cdot U^{T}$ if $\psi=\psi^{T}$).
\item Interpret the index structure of $U$ as $U^{i,j}=U^{i,\alpha_{k}}=U^{\left(s_{1}\cdots s_{k}\right),\alpha_{k}}$.
\item $L^{\alpha'_{k}\mu_{k}\alpha_{k}}=U{}^{*\, s'_{1}\cdots s'_{k},\alpha'_{k}}\cdot H_{s_{1}\cdots s_{k}}^{s'_{1}\cdots s'_{k},\mu_{k}}\cdot U^{s_{1}\cdots s_{k},\alpha_{k}}$\nobreak,
$H_{s_{1}\cdots s_{k}}^{s'_{1}\cdots s'_{k},\mu_{k}}=H_{s'_{1}s_{1}}^{[L]\,\mu_{1}}\cdot H_{s'_{2}s_{2}}^{\mu_{1}\mu_{2}}\cdots H_{s'_{k}s_{k}}^{\mu_{k-1}\mu_{k}}$
\eqref{eq:TransMPO}.
\item Use $V$ analog to $U$ to calculate $R^{\alpha'_{r}\mu_{r}\alpha_{r}}$.
\end{enumerate}
Comparing with \eqref{eq:Q-MPS} we find that
\begin{eqnarray*}
U & = & Q_{[L]\, s_{1}}^{[1]\,\alpha_{1}}\cdots Q_{[L]\, s_{k}}^{[k]\,\alpha_{k-1}\widetilde{\alpha}_{k}}\\
D & = & \lambda^{\widetilde{\alpha}_{k}\alpha_{k}}\\
V & = & Q_{[R]\, s_{k+1}}^{[k+1]\,\alpha_{k}\alpha_{k+1}}\cdots Q_{[R]\, s_{n}}^{[n]\,\alpha_{n-1}}
\end{eqnarray*}

\section{Broken translational invariance \label{sec:broken-translational-invariance}}

In this section, we present some new concepts for the iMPS algorithm
which arise from the need to deal with spontaneously broken translational
invariance. A principal shortcoming of the basic iMPS algorithm is
its failure to converge in such cases. To overcome this deficiency,
we introduce the superposed multi-optimization (SMO) method in section
\ref{sub:Parallel-optimization}. Once convergence is restored, we
turn our attention in section \ref{sub:Which ground state} to the
question of how to obtain a specific solution out of the ground state
manifold degenerate due to broken translational invariance. In addition,
in section \ref{sub:Translational-invariant-ground-1}, we treat the
special case of a non-degenerate ground state which is separated by
a very small energy gap from a state that breaks translational invariance.

\subsection{Preliminary considerations}

Just one tensor $A_{[\textrm{converged}]}$ suffices to construct
an entire iMPS. At first sight, one might therefore think that such
an iMPS is only capable of describing states where all sites behave
equally, which is no longer true for states which break translational
invariance. But still, also these states can be handled. This is due
to the construction rules presented in section \ref{sub:Constructing-the-MPS}.
These result in an iMPS structure given by equation \eqref{eq:Q-MPS},
where the matrix $\lambda$ marks a special position (and with that
breaks translational invariance) if it can not be commuted to its
neighbor sites.

The real problem is to \emph{find} $A_{[\textrm{converged}]}$. In
\ref{sub:The-idea-behind}, we already mentioned that our argumentation
in favor of the convergence $A_{[n]}\rightarrow A_{[\textrm{converged}]}$
is no longer valid in the case of broken translational invariance.
The iMPS algorithm is grounded on local optimization and therefore
it is vulnerable to locally altering states, as they appear on a physical
level for states with broken translational invariance. In this case,
simple local optimization will not result in a global optimal fixed
point $A_{[\textrm{converged}]}$.

\subsubsection{Known solutions}

When we write about the breakdown of translational invariance, we mean
that the state is no longer invariant under the shift of one site.
Still, the state can maintain invariance under the shift of $k$ sites.
If $k$ is known, one can introduce new super sites where one super
site encompasses $k$ of the old sites. Now, the system is translational
invariant with respect to the shift of one super site. This involves
that we have to optimize tensors which represent the super sites.
Due to the exponential increase of the physical dimension, this method
is practical for very small $k$ only. To avoid the scaling problem,
Crosswhite \cite{Crosswhite2008} suggested to use an MPS ansatz for
the super sites. In practice this means, we insert $k$ old sites at
once and optimize the corresponding tensors. We are not aware if this
was ever tested successfully. Aside from that, one still needs a prior
knowledge of the value $k$.

Alternatively, one can extend the standard MPS structure with auxiliary tensors \cite{Ueda2011, Ueda2012},
which allow us to introduce a symmetry breaking element for the price of a non-linear optimization.
To our knowledge, this has not been tested for long-range interactions, so far.

\subsection{Superposed multi-optimization (SMO) \label{sub:Parallel-optimization}}

Our solution of the convergence problem induced by locally altering
states does not depend on any prior knowledge and stays within the standard MPS framework. The key idea is to
wash out local dependency of the optimization by choosing each new
tensor $A_{[n]}$ such that it minimizes the sum of the energy of
exponentially many different MPS instead of just one. These MPS are
different ground state approximations to the qualitatively same Hamiltonian
applied to systems of different sizes. All necessary minimizations
can be joined in a superposition and solved by one optimization. The
time relevant steps stay the same as in the single MPS optimizing
algorithm presented so far, so there is no noticeable loss in speed.

As explained in the following, after each optimization round, the number of MPS
joined in the superposition increases by a factor of 4. Thus, in the $n$th
round, the optimization \eqref{eq:Effektive-Min} gets formally extended
to

\begin{align}
\min\left(\sum_{i=1}^{4^{n-1}}\langle\psi_{i}|H_{i}|\psi_{i}\rangle\right) & \rightarrow\min\left(\sum_{i=1}^{4^{n-1}}\langle A_{[n]}|\mathbb{H}_{[n]\, i}|A_{[n]}\rangle\right)\nonumber \\
 & =\min\left(\langle A_{[n]}|\sum_{i=1}^{4^{n-1}}\mathbb{H}_{[n]\, i}|A_{[n]}\rangle\right)\nonumber \\
 & =\min\left(\langle A_{[n]}|\widetilde{\mathbb{H}}_{[n]}|A_{[n]}\rangle\right).\label{eq:Sum Min}
\end{align}
The MPS representing the $|\psi_{i}\rangle$ are of different length
and the position of the tensor $A_{[n]}$ is no longer in the center
but varies from MPS to MPS. In the basic algorithm, the tensors $A_{[n]}$
experience quite individualized environments and the optimization
adapts to these local circumstances. In the modified algorithm, $A_{[n]}$
faces exponentially many different environments averaging out local
effects and emphasizing common global features. This enforces heavily
the desired convergence $A_{[n]}\rightarrow A_{[\textrm{converged}]}$.

\subsubsection{Modification of the algorithm}

Formally, the superposition $\widetilde{\mathbb{H}}_{[n]}=\sum_{i=1}^{4^{n-1}}\mathbb{H}_{[n]\, i}$
in equation \eqref{eq:Sum Min} is based on $2^{2(n-1)}=4^{n-1}$
different MPS. These MPS are not hand picked, but indirectly generated
by the algorithm. The only modification needed to create such a superposition
concerns the left and right halves $L^{\alpha'_{l}\mu_{l}\alpha_{l}}$
and $R^{\alpha'_{r}\mu_{r}\alpha_{r}}$. We still use equation  \eqref{eq:New Left-Half}
and \eqref{eq:New Right-Half} to perform the iteration steps $L_{[n-1]}^{\alpha'_{l}\mu_{l}\alpha_{l}}\rightarrow L_{[n]}^{\alpha'_{l}\mu_{l}\alpha_{l}}$
and $R_{[n-1]}^{\alpha'_{r}\mu_{r}\alpha_{r}}\rightarrow R_{[n]}^{\alpha'_{r}\mu_{r}\alpha_{r}}$
but afterwards we add the new and the old results to achieve superpositions
\begin{eqnarray}
L_{[n]}^{\alpha'_{l}\mu_{l}\alpha_{l}} & \leftarrow & L_{[n-1]}^{\alpha'_{l}\mu_{l}\alpha_{l}}+L_{[n]}^{\alpha'_{l}\mu_{l}\alpha_{l}}\nonumber \\
R_{[n]}^{\alpha'_{r}\mu_{r}\alpha_{r}} & \leftarrow & R_{[n-1]}^{\alpha'_{r}\mu_{r}\alpha_{r}}+R_{[n]}^{\alpha'_{r}\mu_{r}\alpha_{r}},\label{eq:New Lef Right Half}
\end{eqnarray}
Since this is a simple addition of two tensors, the structure and
size of $L_{[n]}^{\alpha'_{l}\mu_{l}\alpha_{l}}$ and $R_{[n]}^{\alpha'_{r}\mu_{r}\alpha_{r}}$
stay the same and do not entail any computational complications.

In the basic algorithm we have to decide at each iteration step whether
to absorb the tensor $A_{[n-1]}$ into the left half $L_{[n]}^{\alpha'_{l}\mu_{l}\alpha_{l}}$
\eqref{eq:New Left-Half} or into the right half $R_{[n]}^{\alpha'_{r}\mu_{r}\alpha_{r}}$
\eqref{eq:New Right-Half}. Only the half of choice is modified.
Now, we symmetrize the algorithm and modify both halves in each step.
With this modification, the operator sum $\widetilde{\mathbb{H}}_{[n]}$
is calculated with one single use of equation  \eqref{eq:Effektive H}.
As a further advantage of the symmetrization the modified iMPS algorithm
can now take advantage of mirror symmetric Hamiltonians. As this
subject is a bit off-topic, we refer the interested reader to Appendix
\ref{sub:Appendix:-Mirror-symmetry}.

Since each of the tensors $A_{[1]},\dots,A_{[n-1]}$ is absorbed into the left half $L$ and the right half $R$ the longest MPS encoded in the operator superposition of $\widetilde{\mathbb{H}}_{[n]}$ contains $2(n-1)$ tensors (where we neglect the contribution from the initialization routine explained in section \ref{sub:Initialization} and the hole for the new tensor $A_{[n]}$). But in the general MPS each of the $2(n-1)$ tensors  only appears with a probability of $50\%$ due to the addition of the new and old $L$ and $R$ in equation \eqref{eq:New Lef Right Half}.  The possible tensor combinations give rise to the heralded $2^{2(n-1)}$ different MPS  from which $\binom{k}{2(n-1)}$ are of length $k$. More precisely $\binom{k_{[L]}}{(n-1)}\cdot\binom{k_{[R]}}{(n-1)}$ of these MPS contain $k_{[L]}$ tensors left and $k_{[R]}$ tensors right from the hole for the new tensor $A_{[n]}$.

\subsubsection{Comments\label{sub:Comments}}

The crucial observation is that the iteration steps to perform are
always the same independent of the tensor position and the size of
the MPS. Therefore all the different MPS can be optimized together
combined in a superposition. The reader who is more familiar with
finite MPS calculations might wonder about the complete loss of information
concerning the single MPS in the superposition, which comes along
with equation  \eqref{eq:New Lef Right Half}. We have to remind ourselves
that the main objective of the iMPS algorithm is to get the tensor
$A_{[\textrm{converged}]}$, which suffices to construct the infinite
MPS. The finite MPS are just tools to obtain this tensor. Once we
have it the finite MPS are no longer needed.

Another interesting question is whether the operator sum $\widetilde{\mathbb{H}}_{[n]}$  in equation  \eqref{eq:Sum Min} constructed
via equation \eqref{eq:New Lef Right Half} is really suitable for a variational ansatz.
This question has two aspects.
\begin{enumerate}
\item	Is every optimal iMPS solution of equation \eqref{eq:Sum Min} also a minimum of the Hamiltionian?
\item	Does the algorithm always converge to this optimal solution?
\end{enumerate}
The intuitive argumentation given in section \ref{sub:The-idea-behind} also suggests that the answer to the second question should be yes, but actually even for the well-established DMRG algorithm the answer has to be no, since otherwise NP hard problems could be solved. Nonetheless, it is a matter of fact that DMRG converges extremely well for most practical purposes. The same "practical proof" can be given for equation \eqref{eq:Sum Min} as demonstrated by our applications (section \ref{sec:Applications}).

The first question can be answered more formally. Equation \eqref{eq:Sum Min} represents the sum of exponentially many energy terms. The lowest conceivable value of this sum is reached, if each energy term takes its individual minimal value. If all individual energies are minimized, the obvious answer to the first question is yes. Hence, we have to ask, whether it is possible to minimize all individual energies at once, having in mind that all MPS involved are created indirectly via equation \eqref{eq:New Lef Right Half}. With finite MPS this might only be possible up to a certain relative error. But for the limit of infinite MPS this relative error shrinks to zero and the problem is trivially solved by uniform iMPS, i.e.,\ in the case where all tensors stem from the same $A_{\rm [converged]}$. Then, all iMPS appearing in equation \eqref{eq:Sum Min} look alike and either none or all of them minimize their Hamiltonians. Even in the case of broken translational invariance, one can always find at least one translational invariant ground state, which can be written as uniform iMPS and hence optimizes all terms in the sum at once.

Next, we like to further inspect the numerical consequences of equation \eqref{eq:New Lef Right Half}
for the different MPS which are part of the operator sum $\widetilde{\mathbb{H}}_{[n]}$.
The tensor $A_{[n-1]}$ was absorbed into exactly half of the superpositions
encoded in $L_{[n]}$ and in $R_{[n]}$. Since $L_{[n]}$ and $R_{[n]}$
are the building blocks of $\widetilde{\mathbb{H}}_{[n]}$ \eqref{eq:Effektive H},
four subsets of $\widetilde{\mathbb{H}}_{[n]}$ can be distinguished:
\begin{enumerate}
\item $A_{[n-1]}$ was neither absorbed into $L_{[n]}$ nor into $R_{[n]}.$
\item $A_{[n-1]}$ was only absorbed into $L_{[n]}.$
\item $A_{[n-1]}$ was only absorbed into $R_{[n]}.$
\item $A_{[n-1]}$ was absorbed into both halves $L_{[n]}$ and $R_{[n]}$.
\end{enumerate}
Although we never experienced any practical problems, the cases 1
and 4  are, at least from the theoretical point of view, a bit troublesome.
In case 4  the tensor $A_{[n-1]}$ is inserted twice. But $A_{[n-1]}$
was never optimized for double insertion. Close to the end, when $A_{[n]}\rightarrow A_{[\textrm{converged}]}$
is almost achieved, this should pose no problem. Meanwhile, at an early
stage the effect should be more severe. On the other hand, even in
the basic algorithm, the MPS description is not perfect \--- especially
not at the beginning.

Case 1 might seem trivial, since everything stays the same. Potential
difficulties arise in the superposition with the other cases. According
to equation \eqref{eq:Decomposition} the tensors $A$ are decomposed
into $Q$ and $\lambda$ and only $Q$ is absorbed. The matrix $\lambda$
is overwritten with the next $A$ (respectively $Q$). In the basic
algorithm, this is easy to justify: The next $A_{[n]}$ can compensate
for $\lambda_{[n-1]}$. But, a perfect compensation can only be achieved
for \emph{one }$\lambda$, not for many of them. Here is the problem:
In case 1, old $\lambda_{[n-2]},\lambda_{[n-3],}\dots$ of the previous
steps are conserved, while in 2., 3.\  and 4.\  a new $\lambda_{[n-1]}$
comes into play. All $\lambda$ have to be compensated for. The stronger
the $\lambda$ alter, the less adequate is their compensation. The
variation of the $\lambda$ can be reduced by enforcing $\Vert A_{[n]}-A_{[n-1]}\Vert$
to be small. Towards the end of the optimization $\Vert A_{[n]}-A_{[n-1]}\Vert$
is small anyway. At an early stage one might have to resort more strongly
to the convergence enforcing method we will discuss in section \ref{sub:Gain-function}.

\subsection{Selecting a specific ground state\label{sub:Which ground state}}

We have seen how to ensure convergence in the case of broken translational
symmetry. But so far, we have no control to which of the degenerate
ground states the algorithm converges. Some of these ground states
might be more favorable for our purposes than others, and we now answer
the question as to how to obtain them. Any further degeneration aside from
broken translational invariance is excluded from this consideration.

Let us look at two fully converged MPS $\mathcal{A}$ and $\mathcal{B}$
where $\mathcal{B}$ be a representation of the wished-for ground
state which fits our purposes best, while $\mathcal{A}$ stands for
any ground state to which the algorithm actually has converged. According
to equation \eqref{eq:Q-MPS}, both MPS have the following structure
\begin{eqnarray}
\mathcal{A} & = & \cdots Q_{[L]\, s_{-1}}^{\alpha_{-2}\alpha_{-1}}\cdot Q_{[L]\, s_{0}}^{\alpha_{-1}\widetilde{\alpha}_{0}}\cdot\mbox{\ensuremath{\lambda^{\tilde{\alpha}_{0}\alpha_{0}}}}\cdot Q_{[R]\, s_{1}}^{\alpha_{0}\alpha_{1}}\cdot Q_{[R]\, s_{2}}^{\alpha_{1}\alpha_{2}}\cdots\nonumber \\
\mathcal{B} & = & \cdots q_{[L]\, s_{-1}}^{\alpha_{-2}\alpha_{-1}}\cdot q_{[L]\, s_{0}}^{\alpha_{-1}\widetilde{\alpha}_{0}}\cdot\xi^{\tilde{\alpha}_{0}\alpha_{0}}\cdot q_{[R]\, s_{1}}^{\alpha_{0}\alpha_{1}}\cdot q_{[R]\, s_{2}}^{\alpha_{1}\alpha_{2}}\cdots\label{eq:MPS A B im Haupttext}
\end{eqnarray}
In Appendix \ref{sub:Appendix:-Transformation-poof}, we show that
it suffices to replace the matrix $\lambda^{\tilde{\alpha}_{0}\alpha_{0}}$
in $\mathcal{A}$ by the new matrix $\gamma^{\tilde{\alpha}_{0}\alpha_{0}}$
to obtain an MPS which represents exactly the same physical state
as $\mathcal{B}$
\begin{equation}
\mathcal{B}=\cdots Q_{[L]\, s_{-1}}^{\alpha_{-2}\alpha_{-1}}\cdot Q_{[L]\, s_{0}}^{\alpha_{-1}\widetilde{\alpha}_{0}}\cdot\gamma^{\tilde{\alpha}_{0}\alpha_{0}}\cdot Q_{[R]\, s_{1}}^{\alpha_{0}\alpha_{1}}\cdot Q_{[R]\, s_{2}}^{\alpha_{1}\alpha_{2}}\cdots\label{eq:transformation Formel}
\end{equation}
In other words, we do not need to take care to which ground state
the algorithm converges, since after it has converged, we are able
to transform the obtained solution easily into any other. We do not
even have to know $\mathcal{B}$, as long as we have a description
such as, e.g.,\  ``the ground state with the highest expectation value
for the operator $\widehat{X}$''. All we have to do is a one-time
optimization of the new matrix $\gamma^{\tilde{\alpha}_{0}\alpha_{0}}$
under the desired side condition.

\subsubsection{One tensor update versus multi tensor update\label{sub:Bond Dimension}}
So far, we focused on uniform MPS which result from an algorithm that inserts one new tensor each round. Crosswhite \cite{Crosswhite2008} suggested that one might also insert a certain number of $q$ tensors per round in the form of a small MPS. Although we are so far not aware of any successful practical applications of this ansatz, it is worth having a closer look.
For the single-site algorithm to work in the presence of broken translational, invariance we introduced the SMO method, which washes out local variations and thereby fortifies convergence. Still, the SMO is a general method and could also be implemented in an $q$-site algorithm.

As we have seen in the section above, the single-site iMPS algorithm will come up with a solution
that encodes all possible ground states. This abundance has its price. Given the situation that we know the periodicity $q$ of the ground state of interest, we could use an iMPS algorithm which inserts $q$ sites at once. This would enable us to find some specific lowly entangled ground states which could be expressed by a non-uniform MPS with a far smaller bond dimension. Since $q$ translationally shifted copies of such a non-uniform MPS always allow us to construct a uniform MPS, the maximal gain in bond dimension  is given by a factor $q$ and the maximal difference in the entanglement entropy of the half chain is  $\Delta S =\log_2(q)$. We could confirm this difference for the model studied in the applications (section \ref{sec:Applications}) varying the matrix $\lambda$ \eqref{eq:MPS A B im Haupttext} over the set of ground states, as described in the section above.

From the perspective of the needed bond dimension, the multi-tensor update is clearly superior to the-single tensor update for systems with a periodicity $q>1$. Sill, in this paper we favor the single-tensor update, which needs no prior knowledge of the periodicity and results in a well-converging algorithm, which has proofed its reliability in practical tests.

\subsubsection{Degenerate tensor\label{sub:degenerate-tensor}}

In the case of broken translational, invariance one can jump from one
ground state solution to another just by changing the matrix $\lambda$,
which is part of the bigger tensor $A_{[n]}$ \eqref{eq:Decomposition}.
Hence, different $A_{[n]}$ minimize $\langle A_{[n]}|\mathbb{\widetilde{H}}_{[n]}|A_{[n]}\rangle$,
i.e.,\ $A_{[n]}$ is degenerate. The iMPS algorithm aims for the
convergence $A_{[n]}\rightarrow A_{[\textrm{converged}]}$. Without
precautions, this convergence might be undermined towards the very
end by an $A_{[n]}$ which jumps from one solution to another. At
a first glance, this does not seem troublesome, because all solutions
$A_{[n]}$ could jump to are good solutions. Nonetheless, due to imperfect
numerics this jumping might also occur into $A_{[n]}$ of minor quality.
This effect is not fatal, but it still might turn an otherwise perfect
result into a less accurate one.

To suppress this effect, we can resort to the convergence enforcing
method we will present in \ref{sub:Gain-function}. In addition, we
will describe in \ref{sub:Translational-invariant-ground-1} and \ref{sub:Altered-minimization-routine}
a numerical method to enforce a translational invariant solution which
eliminates the above-mentioned degeneration.

\subsection{Translational invariant ground states and local minima\label{sub:Translational-invariant-ground-1}}

In this section, we consider possible convergence problems due to
translational invariance breaking states which lie closely above the
non-degenerate ground state level. In such cases, the infinite system
still provides a translational invariant ground state, while for finite
systems even small alterations of the energy spectrum due to boundary
effects suffice to favor a ground state with broken translational
invariance. Since the iMPS algorithm is based on growing finite systems,
it might start out converging into a false minimum and get trapped
there. Even if the algorithm escapes out of this trap later, it supposably
costs many optimization rounds and significantly slows down convergence.

To avoid these problems we suggest to modify the algorithm such that
it only converges to translational invariant states. This is no limitation:
In the case of an unique ground state, the sole solution has to be
translational invariant, anyway. If the ground state level is degenerate,
one of the solutions is translational invariant and according to equation
\eqref{eq:transformation Formel} we can still transform it into another
type of solution after the algorithm has converged.

Whether the fully converged MPS $\mathcal{A}$,
\begin{equation}
\mathcal{A=}\cdots Q_{[L]\, s_{-1}}^{\alpha_{-2}\alpha_{-1}}\cdot Q_{[L]\, s_{0}}^{\alpha_{-1}\widetilde{\alpha}_{0}}\cdot\mbox{\ensuremath{\lambda^{\tilde{\alpha}_{0}\alpha_{0}}}}\cdot Q_{[R]\, s_{1}}^{\alpha_{0}\alpha_{1}}\cdot Q_{[R]\, s_{2}}^{\alpha_{1}\alpha_{2}}\cdots
\end{equation}
 is translational invariant or not depends on its matrix $\lambda$.
At this point we should be more precise and write $\lambda_{[L]}$
or $\lambda_{[R]}$, depending on whether $\lambda$ stems from a
left or a right decomposition \eqref{eq:Decomposition}. Actually,
as a consequence of decomposition \eqref{eq:Decomposition} the MPS
$\mathcal{A}$ is translational invariant if the left and right version
of $\lambda$ are identical
\begin{equation}
\lambda_{[L]}=\lambda_{[R]}=\lambda\Rightarrow Q_{[L]}\cdot\lambda=A_{[\textrm{converged}]}=\lambda\cdot Q_{[R]}.\label{eq:Lambda L gleich R}
\end{equation}
In this case $\lambda$ can be commuted to any position and hence
does no longer mark any specific site of the MPS. This is what we
are aiming for.

In order to end up with an $A_{[\textrm{converged}]}$ where $\lambda_{[L]}=\lambda_{[R]}$
we alter the minimization routine which computes the tensors $A$
such that solutions with small differences $\Vert\lambda_{[L]}-\lambda_{[R]}\Vert$
i.e.\  big overlap $\langle\lambda_{[L]}|\lambda_{[R]}\rangle$ are
preferred. In the long run this should accumulate to $\lambda_{[L]}=\lambda_{[R]}$.

As a first straight forward way we tried to extend the minimization
\eqref{eq:Sum Min} of $\langle A|\mathbb{\widetilde{H}}|A\rangle$
to
\begin{equation}
\min\left(\langle A|\mathbb{\widetilde{H}}|A\rangle-\gamma_{[\lambda]}\cdot\langle\lambda_{[L]}|\lambda_{[R]}\rangle\right)\label{eq:Nebenbedingung Trans}
\end{equation}
 with a suitable coupling parameter $\gamma_{[\lambda]}$. This is
no longer a simple to solve bilinear problem since one needs to perform
the decomposition \eqref{eq:SVD} to get $\lambda_{[L]}$ and $\lambda_{[R]}$.
To avoid this complication and restore bilinearity we tried to resort
to the easily calculated approximations $\bar{\lambda}_{[L]}$ and
$\bar{\lambda}_{[R]}$ \eqref{eq:Approx Lambda} derived in the appendix
\ref{sub:Appendix:-Altered-minimization}, but the results we obtained
in this way were not very convincing.

In section \ref{sub:Altered-minimization-routine} we introduce a
less conventional approach which turned out to work far more satisfyingly
for us. Instead of extending the minimization of $\langle A|\mathbb{\widetilde{H}}|A\rangle$
by a new term as suggested in equation \eqref{eq:Nebenbedingung Trans},
we alter the routines of the iterative eigenvector solver we use to
solve it. The modus operandi of these solvers is reviewed in section
\ref{sub:Minimization-routine-and}. Until after then, we suspend further
explanations.

\section{Enhanced algorithm\label{sec:Enhanced-algorithm}}

The considerations of the last section were mainly conceptual. The
only actual change of the algorithm we performed is given by equation
\eqref{eq:New Lef Right Half}, which incorporates the SMO method.
In this section, we delve far more into numerical details and extend
the algorithm by further routines to make it more efficient. A reader
not interested in technical details of the algorithm might proceed
directly to section \ref{sec:Applications}

\subsection{Enforcing convergence\label{sub:Enforcing-convergence}}

The goal of the iMPS algorithm is the global convergence $A_{[n]}\rightarrow A_{[\textrm{converged}]}$.
This property has to emerge over the long term, while it is not part
of the evaluation system of the local minimization from which each
$A_{[n]}$ is drawn. As a consequence, small local improvements might
be purchased with strong fluctuating $A_{[n]}$ counteracting global
convergence. In an unstable scenario of overcompensation, these fluctuations
might even inflate in a fatal manner. To prevent this from happening
we extend the algorithm by two methods. The first method (superposition
method) aims at attenuating the influence of problematic $A_{[n]}$
on the ongoing calculations, while the second method (gain function
method) directly modifies the optimization routine such that excessive
variation of the $A_{[n]}$ are suppressed. Both methods are complementary
and worked well together in our calculations.

\subsubsection{Superposition method\label{sub:Superposition-method}}

The first method takes advantage of the fact that the $\mathbb{\widetilde{H}}_{[n]}$
\eqref{eq:Sum Min} of the modified algorithm represent superpositions
of operators. By decreasing the weight of those contributions to the
superpositions which contain problematic $A_{[n]}$, one can ensure
that excessive fluctuation of the $A_{[n]}$ do not spread to the
level of the $\mathbb{\widetilde{H}}_{[n+1]}$ and with that inhibit
a chain of overcompensation. We remind the reader that $A_{[n]}$
is absorbed into $L_{[n+1]}$ \eqref{eq:New Left-Half} and $R_{[n+1]}$
\eqref{eq:New Right-Half} before equation \eqref{eq:New Lef Right Half}
is used to build up superpositions. This latter equation is now replaced
by

\begin{eqnarray}
L_{[n+1]}^{\alpha'_{l}\mu_{l}\alpha_{l}} & \leftarrow & L_{[n]}^{\alpha'_{l}\mu_{l}\alpha_{l}}+\xi_{[n]}\cdot L_{[n+1]}^{\alpha'_{l}\mu_{l}\alpha_{l}}\nonumber \\
R_{[n+1]}^{\alpha'_{r}\mu_{r}\alpha_{r}} & \leftarrow & R_{[n]}^{\alpha'_{r}\mu_{r}\alpha_{r}}+\xi_{[n]}\cdot R_{[n+1]}^{\alpha'_{r}\mu_{r}\alpha_{r}}.\label{eq: Weighted New Left Right Half}
\end{eqnarray}
The only new ingredient compared to equation \eqref{eq:New Lef Right Half}
is the adjustable weight $1\geqslant\xi_{[n]}>0$ calculated as
\begin{equation}
\xi_{[n]}=\min\left(1,\frac{\langle\Delta A\rangle_{[n]}}{\Delta A_{[n]}}\right),\label{eq:Xi-Wert}
\end{equation}
where $\Delta A_{[n]}$ measures the deviation of $A_{[n]}$ and $\langle\Delta A\rangle_{[n]}$
is a weighted average of the previous deviations. Each time the deviation
$\Delta A_{[n]}$ exceeds the average value $\langle\Delta A\rangle_{[n]}$,
$\xi_{[n]}$ gets smaller than 1 and with that the weight of all contributions
of $\mathbb{\widetilde{H}}$ which contain $A_{[n]}$ is reduced accordingly.

To measure the deviation $\Delta A_{[n]}$ we need to define a reference
tensor $A_{[n]}^{[\textrm{refer}]}$ such that $\Delta A_{[n]}=\Vert A_{[n]}-A_{[n]}^{[\textrm{refer}]}\Vert$.
In order to avoid unnecessary fluctuation of this reference tensor
we use the same trick as above and define $A_{[n]}^{[\textrm{refer}]}$
iteratively as a weighted average of the previous $A_{[0\leqslant j<n]}$

\begin{eqnarray}
A_{[n+1]}^{[\textrm{refer}]} & = & \frac{1}{N}\cdot\left(A_{[n]}^{[\textrm{refer}]}+\xi_{[n]}\cdot A_{[n]}\right),\label{eq:InitA}
\end{eqnarray}
with $N=\Vert A_{[n+1]}^{[\textrm{refer}]}\Vert.$ The weights $\xi_{[n]}$
used in equation \eqref{eq:InitA} are the same as in equations \eqref{eq: Weighted New Left Right Half}
and \eqref{eq:Xi-Wert}.

For equation \eqref{eq:Xi-Wert} to work we still have to define the
weighted average $\langle\Delta A\rangle_{[n]}$. Various definitions
are possible. As a heuristic choice, we picked the following one:
\begin{align}
\langle\Delta A\rangle_{[n]}=\frac{1}{N}\cdot\min( & 0.9\cdot\langle\Delta A\rangle_{[n-1]}+0.1\cdot\Delta A{}_{[n-1]},\nonumber \\
 & 1.02\cdot\langle\Delta A\rangle_{[n-1]}),\label{eq:Gleitender Durchschnitt}
\end{align}
with $N=1-0.9^{n}$. Obviously the term $1.02\cdot\langle\Delta A\rangle_{[n]}$
prevents a too sudden increase of $\langle\Delta A\rangle_{[n]}$
by limiting it to $2\%$ per round. Without this term we get the clearer
expression $\langle\Delta A\rangle_{[n]}\sim\sum_{j=0}^{n}0.9^{n-j}\cdot\Delta A_{[j]}$,
i.e.\ older $\Delta A_{[j]}$ lose each round $10\%$ of their influence
in the weighted average.

Finally, we remark that we end up in a deadlock if $\xi_{[n]}=0$.
To prevent this from happening we will introduce the parameter $\Delta_{\textrm{max}}$
in equation \eqref{eq:C_n und D_max} of the upcoming subsection.

\subsubsection{Gain function method\label{sub:Gain-function}}

The idea of the gain function method is to manipulate the minimization
procedure of $\langle A_{[n]}|\mathbb{\widetilde{H}}_{[n]}|A_{[n]}\rangle$
by adding a gain function i.e.\  replacing $\mathbb{\widetilde{H}}_{[n]}$
by $\mathbb{\widetilde{H}}_{[n]}^{[\gamma]}$
\begin{equation}
\mathbb{\widetilde{H}}_{[n]}^{[\gamma]}=\mathbb{\widetilde{H}}_{[n]}-\gamma\cdot|A_{[n]}^{[\textrm{refer}]}\rangle\langle A_{[n]}^{[\textrm{refer}]}|\quad\textrm{with}\quad\gamma\geq0,\label{eq:Convergence H}
\end{equation}
where $|A_{[n]}^{[\textrm{refer}]}\rangle$ is the vectorized version
of the reference tensor defined in equation \eqref{eq:InitA}. Let
$A_{[n]}^{[\gamma]}$ be the result of the above optimization. Obviously,
bigger values for $\gamma$ favor smaller deviations $\Delta A_{[n]}^{[\gamma]}=\Vert A_{[n]}^{[\gamma]}-A_{[n]}^{[\textrm{refer}]}\Vert$.

In the appendix \ref{sub:Appendix:-Gain-function} we show how to
approximate $\gamma$ efficiently such that
\begin{equation}
\Delta A_{[n]}^{[\gamma]}\approx\min\left(c_{[n]}\cdot\Delta A_{[n]}^{[\gamma=0]},\Delta_{\textrm{max}}\right),\label{eq:C_n und D_max}
\end{equation}
where $0<c_{[n]}\leqslant1$ and $\Delta_{\textrm{max}}$ are parameters
of our choice. Limiting $\Delta A_{[n]}^{[\gamma]}$ by assigning
e.g.\  $\Delta_{\textrm{max}}=10\cdot\langle\Delta A\rangle_{[n]}$
\eqref{eq:Gleitender Durchschnitt} ensures that $\xi_{[n]}$ \eqref{eq:Xi-Wert}
is lower bounded around $0.1$ .

Assigning the parameter $0<c_{[n]}\leqslant1$ \eqref{eq:C_n und D_max}
allows us to shorten $\Delta A_{[n]}^{[\gamma]}$ to a chosen fraction
of the maximal value $\Delta A_{[n]}^{[\gamma=0]}$. The price to
pay for a $c_{[n]}<1$ is a lesser energy improvement $\Delta\tilde{E}_{[n]}^{[\gamma]}$
which is calculated as the difference between the energy one gets
due to choosing $A_{[n]}=A_{[n]}^{[\gamma]}$ instead of just taking
$A_{[n]}=A_{[n]}^{[\textrm{refer}]}$

\begin{eqnarray}
\Delta\tilde{E}_{[n]}^{[\gamma]} & = & \langle A_{[n]}^{[\gamma]}|\mathbb{\widetilde{H}}_{[n]}|A_{[n]}^{[\gamma]}\rangle-\langle A_{[n]}^{[\textrm{refer}]}|\mathbb{\widetilde{H}}_{[n]}|A_{[n]}^{[\textrm{refer}]}\rangle.\nonumber \\
 & \thickapprox & \Delta\tilde{E}_{[n]}^{[\gamma=0]}\cdot\left(1-\left(1-c_{[n]}\right)^{2}\right)\label{eq:Energygain pro Schritt}
\end{eqnarray}
Choosing e.g.\  a $\gamma$ which corresponds to $c_{[n]}\thickapprox0.9$
reduces $\Delta A_{[n]}^{[\gamma]}$ by $10\%$ while the energy improvement
$\Delta\tilde{E}_{[n]}^{[\gamma]}$ is still at $99\%$ of the maximal
value $\Delta\tilde{E}_{[n]}^{[\gamma=0]}$.

The parameter $\Delta_{\textrm{max}}$ \eqref{eq:C_n und D_max} and
the entire superposition method are designed to intervene only in
case that $\Delta A_{[n]}$ suddenly increases with ongoing $n$ \---
otherwise they have no effect. The parameter $c_{[n]}$ on the other
hand always effects the calculation if chosen to be smaller than 1.
Generally the $c_{[n]}$ should be chosen in dependence of $\Delta A_{[n]}^{[\gamma=0]}$
(the bigger $\Delta A_{[n]}^{[\gamma=0]}$, the smaller $c_{[n]}$
and vice versa). Just for orientation (not as exclusive choice) we
give the value we chose for most of our calculations

\begin{equation}
c_{[n]}=1-\max\left[0.1;\:0.7+0.1\cdot\log_{10}\left(\Delta A_{[n]}^{[\gamma=0]}\right)\right].
\end{equation}
With that $0.269<c_{[n]}\leqslant0.9$ since $\Delta A_{[n]}^{[\gamma]}\leqslant2$.
This formula was found heuristically and worked fine for us, although
more adequate choices might exist.

When the iMPS algorithm finally approaches its end $\gamma$ becomes
very small and its effect might be overruled by numerical imprecision.
To prevent this, we recommend defining a lower limit for $\gamma$
above the limit of the numerical precision.

\subsection{Energy overgrow\label{sub:Energy-overgrow}}

If the average energy per site of an infinite state does not equal
zero, the total energy of the entire state is $\pm\infty$. Of course,
we never have to deal with an infinite value since our numeric is
restricted to finite systems. Nonetheless, a problem remains. In the
long run, the numeric value of all the information encoded in the
tensor $\mathbb{\widetilde{H}}$ stays more or less the same except
for the energy, which grows with each new site. The tensor $\mathbb{\widetilde{H}}$
gets more and more ill conditioned since the numeric value of the
energy overgrows other information and thereby reduces the achievable
precision. To avoid this problem, we advise to subtract each iteration
step the energy $E_{[n]}=\langle A_{[n]}|\mathbb{\widetilde{H}}_{[n]}|A_{[n]}\rangle$
from the system. Simply speaking, we recommend to assign
\begin{equation}
\mathbb{\widetilde{H}}_{[n+1]}\leftarrow\mathbb{\widetilde{H}}_{[n+1]}-E_{[n]}\cdot\mathbb{I}.
\end{equation}
But to be of any use, this simple assignment has to be encoded into
$L_{[n+1]}^{\alpha'_{l}\mu_{l}\alpha_{l}}$ and $R_{[n+1]}^{\alpha'_{r}\mu_{r}\alpha_{r}}$
the building blocks of $\mathbb{\widetilde{H}}_{[n+1]}$ \eqref{eq:Effektive H}.
This can be done by modifying the MPO tensor $H_{s's}^{\mu_{l}\mu_{r}}$
used in equations \eqref{eq:New Left-Half} and \eqref{eq:New Right-Half}.
As shown in Appendix \ref{sub:Appendix:-MPO-representation},
the MPO tensor $H_{s's}^{\mu_{l}\mu_{r}}$ has a slot which represents
a local interaction term. To this local interaction we add $-E_{[n]}\cdot\mathbb{I}_{s's}$.

\subsection{Minimization routine and information recycling\label{sub:Minimization-routine-and}}

With an increasing number of rounds $n$ the successive minimizations
of the different $\langle A_{[n]}|\mathbb{\widetilde{H}}_{[n]}|A_{[n]}\rangle$
become more and more similar, which opens the opportunity to speed
up the minimization recycling information from preceding turns. In
order to understand these ideas (and also those of section~\ref{sub:Altered-minimization-routine}, section~\ref{sub:LengthOfMPS},
and Appendix~\ref{sub:Subspace-projection-and}), we have to review
the principles of the iterative eigenvector solvers we use \cite{Saad2003}.
In the MPS context, these solvers come with the major advantage that
$\mathbb{\widetilde{H}}_{[n]}$ never has to be constructed explicitly; it suffices to be able to assemble $\mathbb{\widetilde{H}}_{[n]}|A\rangle$
for any given $|A\rangle$. Further, we do not need to perform the
minimization to its very end. For the algorithm to work, it suffices
to perform a limited amount of iterations, such that the resulting
$|A\rangle$ might not be optimal but still significantly improved.
Due to information recycling, these improvements accumulate, such that
the optimal solution emerges in the long run.

Iterative eigenvector solvers are very well suited for the outer eigenvalue
spectrum. Already with modest effort we can expect to find a good
approximation $|e_{0}\rangle\approx|E_{0}\rangle$ for the lowest
eigenvector of $\mathbb{\widetilde{H}}_{[n]}$. The central idea is
to project the problem defined on a huge space of dimension $N$ onto
a much smaller subspace of dimension $k\ll N$ and solve it there.
For this to work, we have to build up iteratively a small set $\left\{ |\mathfrak{A}_{1}\rangle,\dots,|\mathfrak{A}_{k}\rangle\right\} $
of $k$ orthonormal vectors which enables us to express the minimizing
eigenvector $|A_{[n]}^{[\min]}\rangle=|E_{0}\rangle$ of $\mathbb{\widetilde{H}}_{[n]}$
as a linear combination
\begin{equation}
|A_{[n]}^{[\min]}\rangle=|E_{0}\rangle\approx|e_{0}\rangle=\sum_{i=1}^{k}|\mathfrak{A}_{i}\rangle\cdot a_{i}^{[\min]}.\label{eq:Subspace linear combo}
\end{equation}
\begin{eqnarray}
E_{0} & = & \langle E_{0}|\mathbb{\widetilde{H}}_{[n]}|E_{0}\rangle\nonumber \\
 & \approx & a_{i}^{[\min]\,\dagger}\cdot\langle\mathfrak{A}_{i}|\mathbb{\widetilde{H}}_{[n]}|\mathfrak{A}_{j}\rangle\cdot a_{j}^{[\min]}\nonumber \\
 & = & a_{i}^{[\min]\,\dagger}\cdot\mathfrak{H}_{ij}^{[n]}\cdot a_{j}^{[\min]}=e_{0}.\label{eq:Subspace-Matrix}
\end{eqnarray}
We need to solve for $a_{i}^{[\min]}$, which is obviously the minimizing
eigenvector for the $k\times k$ matrix $\mathfrak{H}_{ij}^{[n]}=\langle\mathfrak{A}_{i}|\mathbb{\widetilde{H}}_{[n]}|\mathfrak{A}_{j}\rangle$.
A possible measure for the accuracy of the approximation \eqref{eq:Subspace linear combo}
is given by the norm of the residual vector
\begin{equation}
|r\rangle=\left(\mathbb{\widetilde{H}}_{[n]}-e_{0}\right)|e_{0}\rangle.\label{eq:residual vector}
\end{equation}
As long as $\Vert r\Vert$ is too big we have to extend the set $\left\{ |\mathfrak{A}_{1}\rangle,\dots,|\mathfrak{A}_{k}\rangle\right\} $
iteratively by a further vector $|\mathfrak{A}_{k+1}\rangle$. Any
form of educated guessing for a suitable new $|\mathfrak{A}_{k+1}\rangle$
is allowed. The Lanczos \cite{Lanczos1951} and Arnoldi \cite{Arnoldi1951}
algorithm use a different way of calculation but end up with
\begin{equation}
|\mathfrak{A}_{k+1}\rangle=\Vert r\Vert^{-1}\cdot|r\rangle,\label{eq:Lanczos update}
\end{equation}
with $|\mathfrak{A}_{k+1}\rangle\bot|\mathfrak{A}_{1\leqslant j\leqslant k}\rangle$
by construction.

In contrast to the basic iMPS algorithm, which sets $|\mathfrak{A}_{1}\rangle$
equal to the lowest eigenvector $|e_{[n-1]\,0}\rangle=|A_{[n-1]}\rangle$
of the last round \cite{Crosswhite2008}, we choose
\begin{equation}
|\mathfrak{A}_{1}\rangle=|A_{[n]}^{[\textrm{refer}]}\rangle,\label{eq:First A}
\end{equation}
with $A_{[n]}^{[\textrm{refer}]}$ defined in \eqref{eq:InitA}. This
small change allows an easy implementation of the method presented
in Appendix \ref{sub:Subspace-projection-and} and should help to improve global
convergence. Both versions are straight forward examples of information
recycling since $|A_{[n-1]}\rangle$ as well as $|A_{[n]}^{[\textrm{refer}]}\rangle$
are already good approximations for $|E_{0}\rangle$ (=$|E_{[n]\,0}\rangle)$.
In many cases, we could obtain a considerable speed up extending this
idea to a few more than just the first vector of the set
\begin{eqnarray}
|\mathfrak{A}_{1+j}\rangle & = & |A_{[n-j]}^{[\textrm{refer}]}\rangle_{\bot}\\
 & = & \frac{1}{\Vert\mathfrak{A}_{1+j}\Vert}\cdot\left(\mathbb{I}-\sum_{i=1}^{j-1}|\mathfrak{A}_{i}\rangle\langle\mathfrak{A}_{i}|\right)\cdot|A_{[n-j]}^{[\textrm{refer}]}\rangle.\nonumber
\end{eqnarray}
 Further, we observe that all $|A^{[\textrm{refer}]}\rangle$ \eqref{eq:InitA}
are derived from the best eigenvectors of the previous rounds. As
an additional extension we also tried to include the next best eigenvectors
$|e_{[n-1]\, j>0}\rangle$ of the last round
\begin{equation}
|\mathfrak{A}_{1+j}\rangle=|e_{[n-1]\, j}\rangle,
\end{equation}
The improvements we achieved in this way were relatively poor. A much
more promising way to take advantage of the $|e_{[n-1]\, j}\rangle$
is to use them for an efficient approximation of the inverse operator
$\mathfrak{D}=\left(e_{0}\cdot\mathbb{I}-\mathbb{\widetilde{H}}_{[n]}\right)^{-1}$
\eqref{eq:Davidson Roh-Invers}, which allows a handy implementation
resembling the Davidson (or Jacobi-Davidson) \cite{DavidsonJacobi}
method. As a result the update equation \eqref{eq:Lanczos update}
is replaced by the more appropriate ansatz \eqref{eq:Fast-End-Davidson}.
Details are explained in the Appendix \ref{sub:Appendix:-Davidson-implementation}.

At the end of this section, we like to caution the reader that the
methods presented here might counteract the methods presented in section
\ref{sub:Enforcing-convergence}. Global convergence and improved
local minimization often go hand in hand, but not always. If
the algorithms indicates to run unstable, one should consider to partially
switch off the improvements just presented. This is likely to happen
if $|e_{0}\rangle$ is degenerate. In this case, the recycled knowledge
from the past strongly increases the probability that already a shallow
optimization suffices to find alternative solutions, which might result
in unwanted fluctuations as, e.g.,\  described in \ref{sub:degenerate-tensor}.
Usually, this problem is announced in advance. In the Davidson implementation,
one should not resort to eigenvectors with eigenvalues too close to
the best. Similarly, once the small set of recycled initial values $|A_{[n-j]}^{[\textrm{refer}]}\rangle$
suffices to get a second best eigenvalue very close to the best, it
might be wise to abandon this method and only use $|A_{[n]}^{[\textrm{refer}]}\rangle$
alone.

\subsection{Enforcing translational invariant ground states \label{sub:Altered-minimization-routine}}

In this section, we demonstrate the algorithmic realization of the
considerations put forward in section \ref{sub:Translational-invariant-ground-1}.
There, we argued that it is beneficial to push the algorithm towards
translational invariant iMPS solutions to avoid getting trapped in
local minima. We further showed that translational invariance is assured
if the decomposition \eqref{eq:Decomposition} of the tensor $A_{[\textrm{converged}]}$
results in $\lambda_{[L]}=\lambda_{[R]}$ \eqref{eq:Lambda L gleich R}.
This is what we are aiming for.

The approach we are about to present is not very intuitive. Therefore,
we start our explanations with an intermediate step and introduce
a less practical but easier to understand procedure which consists
of the following steps and has to be performed with each new tensor
$A$ after it has been optimized:
\begin{enumerate}
\item Decompose $A$ into $Q_{[L]}\cdot\lambda_{[L]}=A=\lambda_{[R]}\cdot Q_{[R]}$
\eqref{eq:Decomposition}.
\item Define $\lambda_{[\textrm{sym}]}=\frac{1}{2}\left(\lambda_{[L]}+\lambda_{[R]}\right)$.
\item Set $A\leftarrow\frac{1}{2}\left(Q_{[L]}\cdot\lambda_{[\textrm{sym}]}+\lambda_{[\textrm{sym}]}\cdot Q_{[R]}\right)$.
\item Goto 1.
\end{enumerate}
Due to line 2.\  this procedure converges towards a tensor $A$ with
$\lambda_{[L]}=\lambda_{[R]}$. Further we expect $A^{[\textrm{initial}]}\approx A^{[\textrm{final}]}$
if already $\lambda_{[L]}^{[\textrm{initial}]}\approx\lambda_{[R]}^{[\textrm{initial}]}$.
Nonetheless, the changes in $A$ might be too pronounced to be acceptable.
To soften this approach one can ignore line 4.\  and just go through
1.\ to 3.\ once. After that we generally still have $\lambda_{[L]}\neq\lambda_{[R]}$
but with a reduced distance $\Vert\lambda_{[L]}-\lambda_{[R]}\Vert$
compared to the initial value. This is all we need to achieve $\lambda_{[L]}=\lambda_{[R]}$
in the long run. But the new $A$ is still likely not to qualify for
the optimizing tensor we are looking for.

Now, we come to the procedure we really use. Instead of symmetrizing
the tensor $A$ \emph{after} its optimization we integrate the symmetrization
into the optimization routine. As recapitulated in section \ref{sub:Minimization-routine-and},
the optimization routine expresses the vectorized tensor $A_{[n]}=|A\rangle$
as a linear combination
\begin{equation}
|A\rangle=|\mathfrak{A}_{i}\rangle\cdot a_{i}\label{eq:Nochmal linear combi}
\end{equation}
of a small set of basis vectors $|\mathfrak{A}_{i}\rangle$ \eqref{eq:Subspace linear combo}.
The idea is to alter these basis vectors $|\mathfrak{A}_{i}\rangle$
such that we have a similar effect as the procedure above. At the
stage of the optimization the $Q_{[L/R]}^{[n]}$ are still unknown
and we have to approximate them by their precursors $Q_{[L/R]}^{[n-1]}$.

The $|\mathfrak{A}_{i}\rangle$ are created iteratively. In each iteration
step we first create a new $|\mathfrak{A}_{i}\rangle$ as we used
to do \eqref{eq:Lanczos update}, \eqref{eq:David Referenz} and then
alter it. Therefor we introduce $|\mathfrak{\bar{A}}_{i}\rangle$
defined as

\begin{eqnarray}
|\mathfrak{\bar{A}}_{i}\rangle & = & \frac{1}{2}|\mathfrak{A}_{i}\rangle+\frac{1}{4}\left(Q_{[L]}^{[n-1]}\cdot\bar{\lambda}_{[R]\, i}+\bar{\lambda}_{[L]\, i}\cdot Q_{[R]}^{[n-1]}\right)\nonumber \\
 & \textrm{with} & \bar{\lambda}_{[R]\, i}^{\alpha\beta}=|\mathfrak{A}_{i}\rangle_{s}^{\alpha\gamma}\cdot Q_{[R]\, s}^{[n-1]*\,\gamma\beta}\nonumber \\
 &  & \bar{\lambda}_{[L]\, i}^{[n]\,\alpha\beta}=Q_{[L]\, s}^{[n-1]*\,\alpha\gamma}\cdot|\mathfrak{A}_{i}\rangle_{s}^{\gamma\beta},\label{eq:Neue Trans Inv Basis Vektoren}
\end{eqnarray}
where we tensorized the vector $|\mathfrak{A}_{i}\rangle$ in lines
2 and 3. With that, we replace $|\mathfrak{A}_{i}\rangle$ by an orthonormal
version of $|\mathfrak{\bar{A}}_{i}\rangle$:
\begin{eqnarray}
|\mathfrak{A}_{i}\rangle & \leftarrow & \left(\mathbb{I}-\sum_{j=1}^{i-1}|\mathfrak{A}_{j}\rangle\langle\mathfrak{A}_{j}|\right)\cdot|\mathfrak{\bar{A}}_{i}\rangle\nonumber \\
|\mathfrak{A}_{i}\rangle & \leftarrow & \frac{1}{\Vert\mathfrak{A}_{i}\Vert}\cdot|\mathfrak{A}_{i}\rangle.\label{eq:Neue BasisVektoren}
\end{eqnarray}
For a better understanding we insert the $|\mathfrak{\bar{A}}_{i}\rangle$
in the linear combination \eqref{eq:Nochmal linear combi}. As shown
in Appendix \ref{sub:Appendix:-Altered-minimization}, we get
\begin{eqnarray}
|\bar{A}\rangle & = & |\mathfrak{\bar{A}}_{i}\rangle\cdot a_{i}\nonumber \\
 & \approx & \frac{1}{2}\left(Q_{[L]}\cdot\lambda_{[\textrm{sym}]}+\lambda_{[\textrm{sym}]}\cdot Q_{[R]}\right)\quad\textrm{with}\nonumber \\
\lambda_{[\textrm{sym}]} & = & \frac{1}{2}\left(\lambda_{[L]}+\lambda_{[R]}\right),\label{eq:Symmetrisch A}
\end{eqnarray}
which mimics the effect of the procedure presented above. But in contrast
to the procedure above the story does not end here. The important
point to notice is that the algorithm can still adopt to the alteration
\eqref{eq:Neue BasisVektoren} of the basis vectors $|\mathfrak{A}_{i}\rangle$
and come up with alternative solutions. More favorable weights $a_{i}$
than those used in equation \eqref{eq:Symmetrisch A} are presumably
to be found. Even the $|\mathfrak{A}_{i}\rangle$ themselves are likely
to be different since they are calculated iteratively according to
the needs of the minimization. While there are still enough resources
to compensate sufficiently for the negative effects of the enforced
alteration, the positive effects should survive since the arguments
in their favor are largely independent of the $a_{i}$, $|\mathfrak{A}_{i}\rangle$
chosen by the optimization routine. Still, this alteration is a trade
off, but we have good reasons to believe that we gain more than we
sacrifice.

For practical applications, we only need a few lines of code to implement
equation \eqref{eq:Neue Trans Inv Basis Vektoren}, which is also
easy to turn off for systems where it is not needed, i.e.,\  when the
unaltered algorithm shows no tendency to run the risk of being trapped
in a local minimum. In such a case, the alteration is likely to slow
down the algorithm slightly. For the applications tested by us the
loss in performance was only marginal. On the other hand we also encountered
many cases where the altered algorithm clearly outperformed the
unaltered one, which was partially even unable to find the correct
ground state within the observed run time.

Although we strongly recommend to implement the alteration \eqref{eq:Neue BasisVektoren}
as presented, one could also use a compromise and only alter the first
basis vector $|\mathfrak{A}_{1}\rangle=A_{[n]}^{[\textrm{refer}]}$,
which has already a strong impact on the outcome of the optimization.
This reduced version does not come with the need to program a new
eigenvector solver. Each solver which accepts an initial vector $|\mathfrak{A}_{1}\rangle$
will do. In any case, the gain function in equation \eqref{eq:Convergence H}
is understood to change accordingly to the alteration of $|\mathfrak{A}_{1}\rangle=A_{[n]}^{[\textrm{refer}]}$.

\subsection{Length of the MPS\label{sub:LengthOfMPS}}

After $n$ optimization steps, even the longest MPS encoded in the superposition created by the SMO method
does not surpass the length $l=2\cdot n+l_{0}$ (where $l_{0}$ is the initial length (see \ref{sub:Initialization})).
For some systems with long-range correlations this might be too short unless $n$ reaches
some considerably high number, which would go along with an extended
calculation time. To shorten this calculation time two methods might
be of help:
\begin{enumerate}
\item Use a tensor $A_{[n]\, s}^{\alpha_{l}\alpha_{r}}$ with small bond
dimension $\chi_{[\textrm{small}]}$ until a certain MPS length is
reached, then increase the bond dimension to its final value $\chi_{[\textrm{big}]}$.
\item Use fast Krylov subspace methods \cite{Saad2003} to insert the same
tensor many times (e.g.\  $10^{5}$) into the MPS.
\end{enumerate}
A simple and comfortable way to increase the bond dimension from $\chi_{[\textrm{small}]}$
to $\chi_{[\textrm{big}]}$ is to use an isometric $\chi_{[\textrm{small}]}\times\chi_{[\textrm{big}]}$-matrix
$u^{\alpha\beta}$ with$ $
\begin{equation}
u^{\alpha\beta}\cdot\left(u^{T}\right)^{\beta\alpha'}=\delta^{\alpha\alpha'}
\end{equation}
and proceed as follows after $A_{[n]}$ has been optimized but still
not been inserted into $L_{[n]}$ \eqref{eq:New Left-Half} and $R_{[n]}$
\eqref{eq:New Right-Half}:
\begin{eqnarray}
L_{[n]}^{\beta'_{l}\mu_{l}\beta_{l}} & \leftarrow & L_{[n]}^{\alpha'_{l}\mu_{l}\alpha_{l}}\cdot u^{\alpha'_{l}\beta'_{l}}\cdot u^{\alpha_{l}\beta_{l}}\nonumber \\
R_{[n]}^{\beta'_{r}\mu_{r}\beta_{r}} & \leftarrow & R_{[n]}^{\alpha'_{r}\mu_{r}\alpha_{r}}\cdot u^{\alpha'_{r}\beta'_{r}}\cdot u^{\alpha_{r}\beta_{r}}\nonumber \\
A_{[n]\, s}^{\beta_{l}\beta_{r}} & \leftarrow & A_{[n]\, s}^{\alpha_{l}\alpha_{r}}\cdot u^{\alpha_{l}\beta_{l}}\cdot u^{\alpha_{r}\beta_{r}}.\label{eq:BondDimTransformation}
\end{eqnarray}
Next, the new tensor $A_{[n]\, s}^{\beta_{l}\beta_{r}}$ is inserted
into $L_{[n]}^{\beta'_{l}\mu_{l}\beta_{l}}$ and $R_{[n]}^{\beta'_{r}\mu_{r}\beta_{r}}$
as usual but without the superposition building step \eqref{eq:New Lef Right Half}
respectively \eqref{eq: Weighted New Left Right Half}. To avoid trapping
into a local minimum one might also consider to add a small amount
of noise to $A_{[n]\, s}^{\alpha_{l}\alpha_{r}}$ before applying
equation \eqref{eq:BondDimTransformation}.

A possible strategy for the small bond dimension $\chi_{[\textrm{small}]}$
is to proceed until convergence has been reached $A_{[n]}\rightarrow A_{[\textrm{converged}]}$,
but before the bond dimension is increased, many more copies of $A_{[\textrm{converged}]}$
are inserted into the MPS without any further optimization. These
insertions can be done in the standard fashion or generally much faster
by projecting the problem onto a small subspace, similar to the way
the eigenvector problem is solved (see \ref{sub:Minimization-routine-and}).
To formalize this method, let us introduce the operator $\mathcal{I}$
which inserts one copy of $A_{[\textrm{converged}]}$ into $L_{[n]}$
\eqref{eq:New Left-Half}, i.e.,
\begin{equation}
\mathcal{I}\cdot L_{[n]}=L_{[n+1]}.
\end{equation}
With that we build up the Krylov subspace $\mathcal{K}_{r}$,
\begin{equation}
\mathcal{K}_{r}=\textrm{span}\left\{ L_{[n]},\mathcal{I}\cdot L_{[n]},\mathcal{I}^{2}\cdot L_{[n]},\dots,\mathcal{I}^{r-1}\cdot L_{[n]}\right\}
\end{equation}
and similar with $R_{[n]}$ \eqref{eq:New Right-Half}. As in \ref{sub:Minimization-routine-and}
we create an orthonormalized system of basis vectors $|\mathfrak{L}_{k}\rangle$
\begin{equation}
|\mathfrak{L}_{k}\rangle=\frac{1}{\Vert\mathfrak{L}_{k}\Vert}\cdot\left(\mathbb{I}-\sum_{i=0}^{k-1}|\mathfrak{L}_{i}\rangle\langle\mathfrak{L}_{i}|\right)\cdot|\mathcal{I}^{k}\cdot L_{[n]}\rangle
\end{equation}
 and calculate $\mathfrak{I}_{ij}$ the subspace projection of $\mathcal{I}$
\begin{equation}
\mathfrak{I}_{ij}=\langle\mathfrak{L}_{i}|\mathcal{I}|\mathfrak{L}_{j}\rangle.
\end{equation}
Keeping in mind that the subspace projection of $L_{[n]}$ is simply
given by the vector $l_{j}=\left(\begin{array}{ccccc}
1 & 0 & 0 & 0 & \ldots\end{array}\right)^{T}$ we find
\begin{equation}
L_{[n+p]}=\mathcal{I}^{p}\cdot L_{[n]}\approx\left(\left(\mathfrak{I}^{p}\right)_{ij}\cdot l_{j}\right)\cdot|\mathfrak{L}_{i}\rangle.\label{eq:Subspace Potenz}
\end{equation}
The number of basis vectors $|\mathfrak{L}_{i}\rangle$ should be
chosen such that this approximation is perfect within computer precision.
Further errors are introduced by an imperfectly converged $A_{[\textrm{converged}]\, s}^{\alpha_{l}\alpha_{r}}$
and from the energy overgrow effect described in \ref{sub:Energy-overgrow}
which should rule out attempts to go for $p\rightarrow\infty$. Still,
a small amount of the last two errors is acceptable since they have
a similar effect as the afore-mentioned extra noise to avoid local
minima.

\begin{figure*}
\includegraphics[width=1\textwidth]{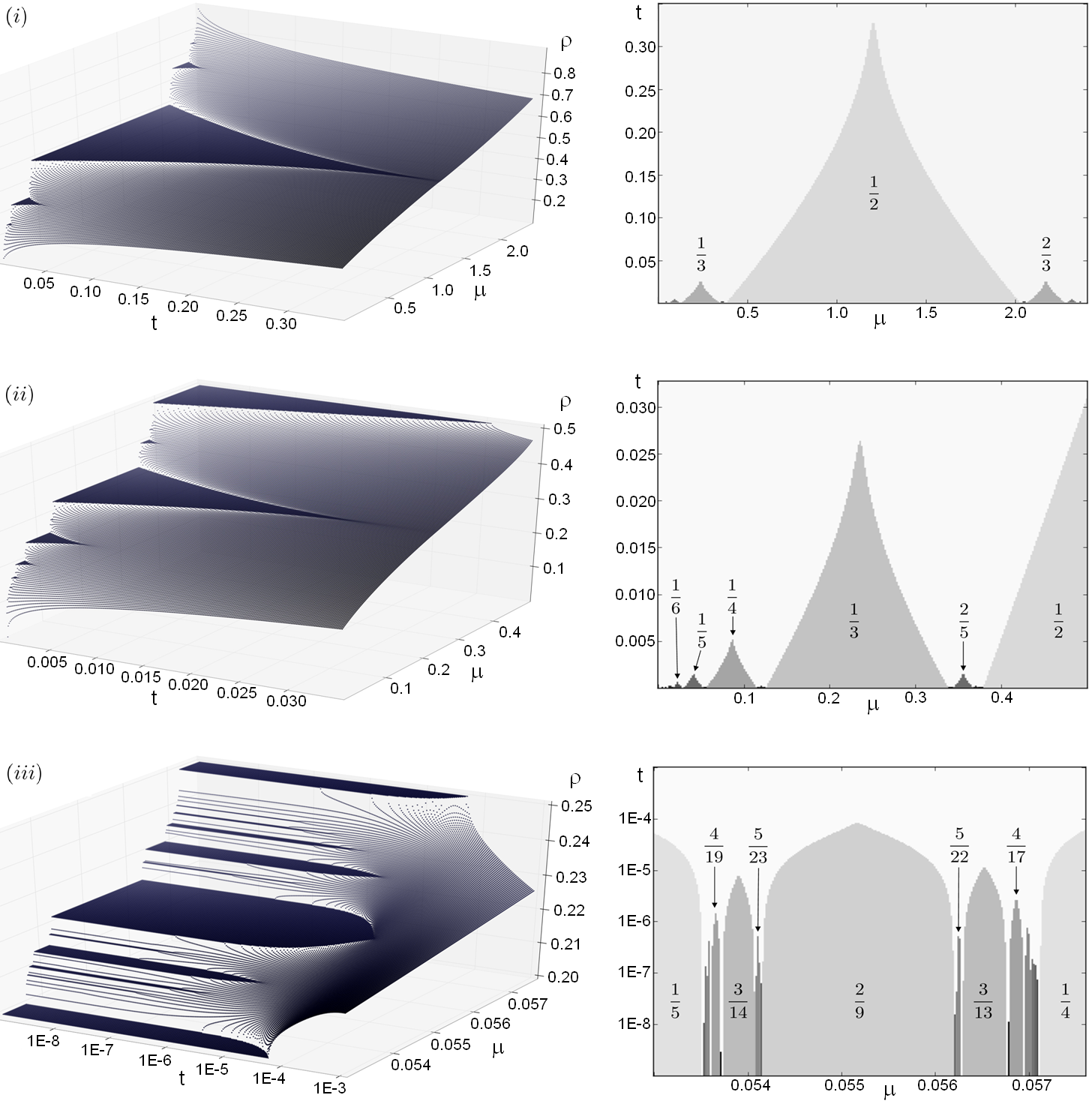}\caption{The densities $\rho$ of polar bosons (left) and the corresponding phase diagrams (right) for the ground states of the Hamiltonian
\eqref{eq:Bose-Hubbard Ham} for $U\rightarrow\infty$ plotted over $t$ and $\mu$ in units of $V=1$.
Each plot consists of 66049 data points calculated with the bond dimensions $\chi_{\textrm{MPS}}=32$ for $(i)$ \& $(ii)$ and $\chi_{\textrm{MPS}}=64$ for $(iii)$.
The fractions associated to selected phases of the phase diagrams denote their $p/q$ values (see main text).\label{fig:SammelPlot}}
\end{figure*}

\begin{figure*}
\includegraphics[width=1\textwidth]{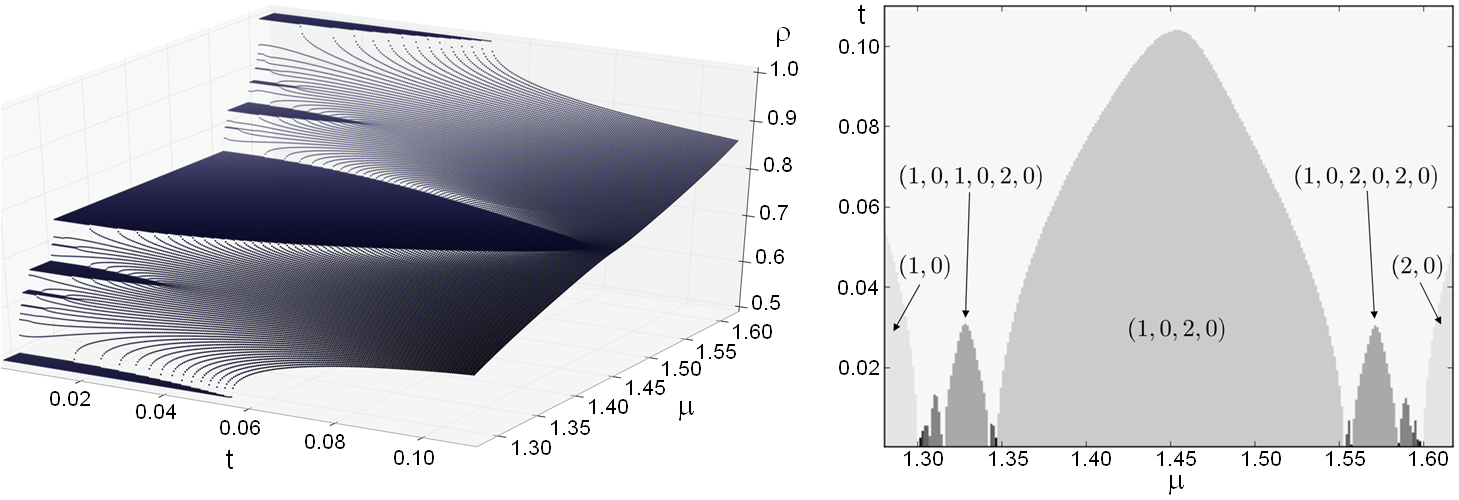}\caption{The densities $\rho$ of polar bosons (left) and the corresponding phase diagram (right) for the ground states of the Hamiltonian
\eqref{eq:Bose-Hubbard Ham} with $U=V=1$ plotted over $t$ and $\mu$.
Each plot consists of 62194 data points calculated with the bond dimensions $\chi_{\textrm{MPS}}=32$.
The bracketed numbers associated to selected phases of the phase diagram describe their periodically repeated occupation pattern (see main text).\label{fig:X0Phase}}
\end{figure*}

\begin{figure*}
\includegraphics[width=1\textwidth]{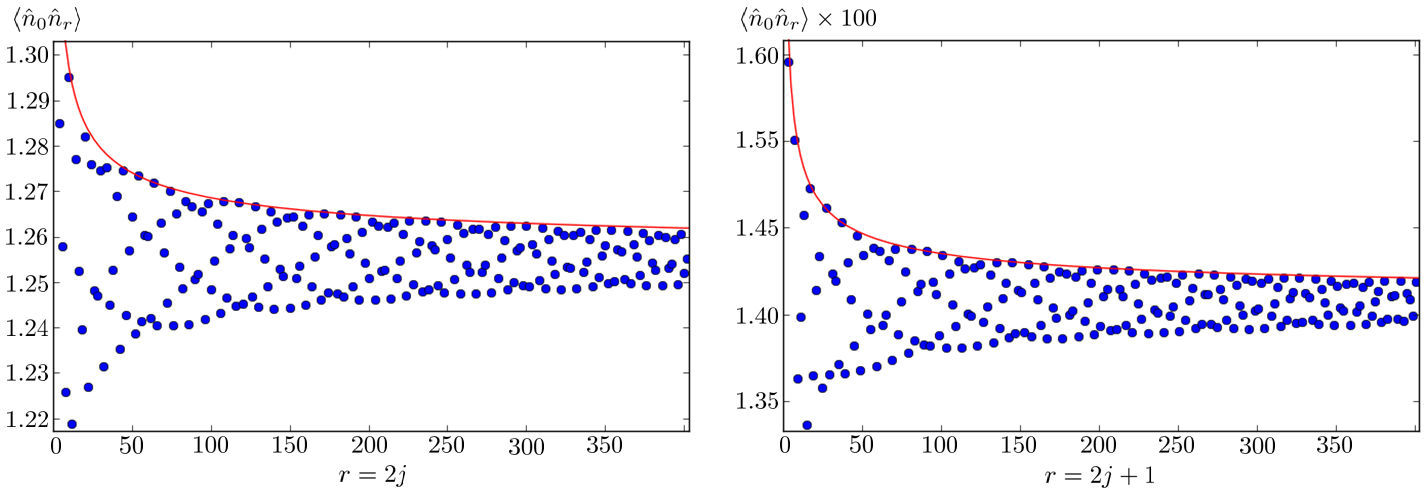}
\caption{The blue circles (color online) mark the values of the correlation function $\langle\hat{n}_{0}\hat{n}_{r}\rangle$ for $U=V=1$, $\mu=1.55$ and $t=0.044$ extracted from an MPS with bond dimension $\chi=256$. For odd distances $r=2j+1$ (right picture) $\langle\hat{n}_{0}\hat{n}_{r}\rangle$ is suppressed by  a factor of roughly 100 compared to even distances $r=2j$ (left picture). To demonstrate the algebraic decay of $\langle\hat{n}_{0}\hat{n}_{r}\rangle$ the function  $f(r) = \alpha\cdot r^{-0.5}+\langle\hat{n}_{0}\hat{n}_{r\rightarrow\infty}\rangle$ with an adequate $\alpha$ is included on top of both plots as a red line.
\label{fig:NN_Korrelation}}
\end{figure*}

\section{Applications\label{sec:Applications}}
In the last section we presented various methods to improve the performance
of the iMPS algorithm with long-range interactions. The main subject
was to ensure convergence, where special attention was paid to broken
translational invariance. This so far troublesome case can now be
tackled mainly due to the newly introduced method of superposed multi-optimization
(SMO).

The modified iMPS algorithm has superior convergence
properties compared to the basic version but it does not surpass its
precision, which is determined by MPS and MPO inherited limitations.
In cases where both versions converge, the quality of the results is
identical. Readers who are interested in the achievable precision of the iMPS method in comparison with analytical solutions
are therefore referred to the literature \cite{Crosswhite2008,McCulloch}.

We checked our algorithm with different models. For a reliable basic benchmark, we examined, e.g.,\ states with long-range chiral order in the next nearest neighbor Heisenberg model and found the expected agreement with the results given in the references  \cite{Furukawa2012,Okunishi2008}.

Here, we will present results for a model of polar bosons described by a Bose-Hubbard like Hamiltonian with a long-range interaction term.
In the thermodynamic limit the ground state of this model exhibits symmetry breaking crystalline phases as well as incommensurate phases with algebraically decaying long-range correlations.

The long-range interaction of the Hamiltonian we consider decays as $\sim r^{-3}$. To model this interaction with an MPO the decay is approximated as weighted sum of $20$
exponential functions
\begin{equation}
(r)^{-3}\approx\sum_{i=1}^{20}a_{i}\cdot\lambda_{i}^{r-1}\quad r=1,2,3,...\label{eq:r-cube0}
\end{equation}
(see also equation \eqref{eq:r-cube} in the appendix and reference~\cite{PolyExp}).

\subsection{Bose-Hubbard model with long-range interaction}

We study the thermodynamic limit ground states of polar bosons in an one-dimensional optical lattice described
by the following effective Hamiltonian \cite{Burnell2009,BurnellAlone}
\begin{eqnarray}
\mathcal{H} & = & V\cdot\sum_{k<j}\frac{1}{(j-k)^{3}}\cdot\hat{n}_{k}\cdot\hat{n}_{j}+\frac{U}{2}\cdot\sum_{j}\hat{n}_{j}\cdot(\hat{n}_{j}-1)\nonumber \\
 &  & -\mu\cdot\sum_{j}\hat{n}_{j}-t\cdot\sum_{j}\left(\hat{c}_{j}^{\dagger}\cdot\hat{c}_{j+1}+\hat{c}_{j}\cdot\hat{c}_{j+1}^{\dagger}\right),\label{eq:Bose-Hubbard Ham}
\end{eqnarray}
where $\hat{c}_{j}^{\dagger}$ and $\hat{c}_{j}$ are the creation
and annihilation operators for a boson on site $j$ and $\hat{n}_{j}=\hat{c}_{j}^{\dagger}\cdot\hat{c}_{j}$.
This model is characterized by a hopping amplitude $t$, an on-site
interaction energy $U$, a chemical potential $\mu$, and a long-range
dipole-dipole coupling $V/r^{3}.$ For $\mu>0$, the chemical potential
favors as many bosons as possible in the ground state, while the dipole-dipole
coupling together with the on-site interaction try to avoid two bosons
coming too close to each other. For certain parameter regimes, this
interplay allows for translational invariance breaking crystalline
phases with optimized distances between the bosons where $q$ sites
accommodate exactly $p$ bosons. We will refer to them as $p/q$-phases.
The model is known to host an entire Devil's Staircase of crystalline
phases for $t=0$ if the joined potential of on-site interaction and
dipole-dipole coupling is convex  \cite{DevilStaircaseOriginal,Burnell2009}.

In the following, we will investigate two qualitative different regimes
of this model: $U\rightarrow\infty$ and $U=V.$

\subsubsection{Devils's Staircase for $U\rightarrow\infty$}

For $U\rightarrow\infty$, each site can accommodate at most one boson
and the effective dimension $d_{\textrm{eff}}$ of the local Hilbertspaces
reduces to $d_{\textrm{eff}}=2$. Figure~\ref{fig:SammelPlot} displays
the average ground state densities of the bosons and the localization
of the corresponding $p/q$-phases.
To determine the periodicity $q$ of the phases we counted the number
of eigenvectors of the transfer matrix $T_{[L]}^{ij}$
\begin{equation}
T_{[L]}^{ij}=T_{[L]}^{\left(\alpha_{l}\alpha'_{l}\right),\left(\alpha_{r}\alpha'_{r}\right)}=Q_{[L]\, s}^{\alpha_{l}\alpha_{r}}\cdot Q_{[L]\, s}^{*\:\alpha'_{l}\alpha'_{r}}
\end{equation}
with an absolute eigenvalue of one. Once $q$ is known $p$ follows
from the average density. Figure \ref{fig:SammelPlot}~$(iii)$ shows a magnification
of the area between the $1/4$-phase and the $1/5$-phase. The biggest phase between these two phases is the $2/9$-phase,
which can be understood as primary compromise ($2/9=\left[1+1\right]/\left[4+5\right]$).
In the same fashion we find e.g.\  that the biggest phase between
the phase $2/9$ and $1/5$ is the $3/14$-phase. The maximal detectable
value of $q$ is given by the bond dimension $\chi$ of the MPS, which
is $64$ in case of Figure \ref{fig:SammelPlot} $(iii)$. However, since the
range of $\mu$ covered by the different phases diminishes with growing
value of $q$ most phases beyond $q=30$ escaped our resolution. The
highest value we hit was $p/q=11/52$.

\begin{figure}
\includegraphics[width=1\columnwidth]{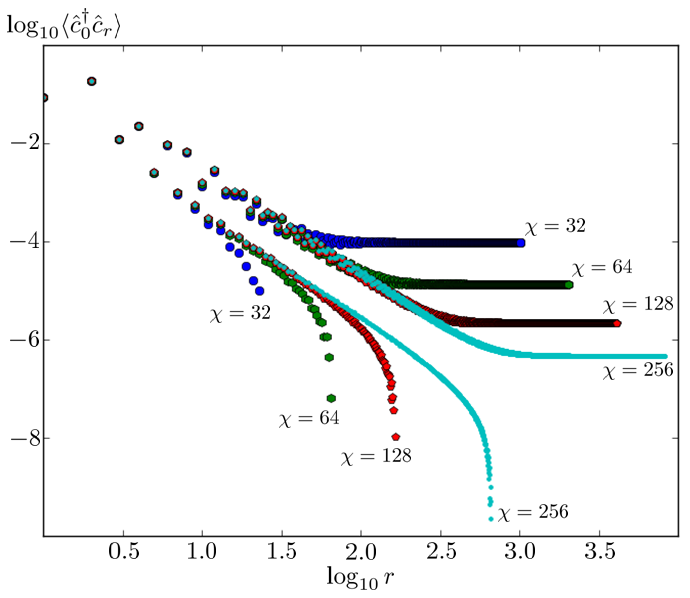}
\caption{Log-log plot of the correlation function $\langle\hat{c}_{0}^{\dagger}\hat{c}_{r}\rangle$ for $U=V=1$, $\mu=1.55$ and $t=0.044$
extracted from MPS with different bond dimensions $\chi$. Due to the spatial order $\langle\hat{c}_{0}^{\dagger}\hat{c}_{r}\rangle$
splits into two branches, each decaying $\propto r^{-\alpha} + {\rm const}_\chi$, with ${\rm const}_\chi<0$ for the lower branch. This constant is a purely numerical effect, as can be seen by the scaling with the bond dimension $\chi$.
\label{fig:CC_Korrelation}}
\end{figure}

\subsubsection{Devils's Staircase for $U=V=1$}

For sufficient small $U$ the ground states of the Hamiltionian~\eqref{eq:Bose-Hubbard Ham} might accommodate more
than one boson per site, which allows for new types of Devil's Staircases. An example is given by
Figure~\ref{fig:X0Phase}, which shows the densities and phases for $U=V=1$. Here, simple translational invariance is broken by an underlying occupation pattern given by $\dots,x_{i},0,x_{i+2},0,x_{i+4},0,\dots$\ with $x_{j}=1\textrm{ or }2$. Of course, for any non-zero hopping amplitude  $t>0$ we expect fluctuations around this pattern such that a more accurate description might be given by  $\dots,x_{i},\varepsilon_{i+1},x_{i+2},\varepsilon_{i+3},\dots$, which we need in the next subsection \ref{sub:Supersolid}. At a certain point, these fluctuations will become so strong that the underlying pattern is destroyed, but this is not the case for the entire region of Figure~\ref{fig:X0Phase}.

In the lobes of the new Devil's Staircase the sublattice $\dots,x_{i},x_{i+2},x_{i+4},\dots$ crystallizes in regular pattern of single and double occupied sites.
These lobes exhibit an approximate symmetry under the exchange of single and double occupied sites. This is a nontrivial symmetry in contrast to
the exact particle hole symmetry of  Figure~\ref{fig:SammelPlot}~$(i)$.

Outside the crystalline phases Burnell \cite{Burnell2009} predicted a supersolid like phase. In the following we present numerical evidence which supports this claim.

\begin{figure}
\includegraphics[width=1\columnwidth]{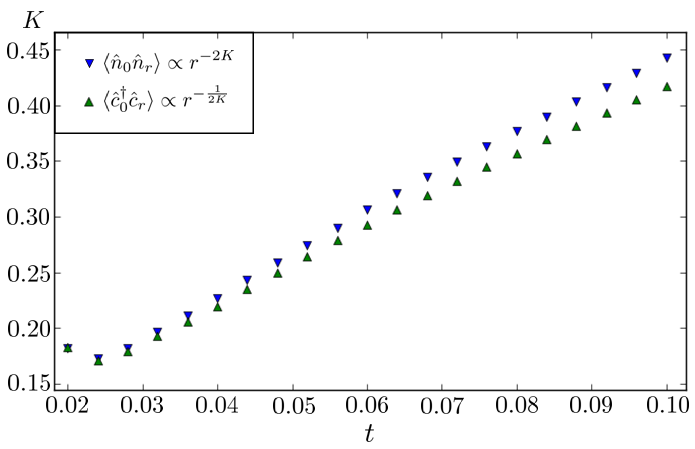}
\caption{The Luttinger liquid parameter $K$ in dependence of the hopping amplitude $t$ for $U=V=1$ and $\mu=1.55$.
The parameter $K$ was obtained in two different ways by fitting the first order term approximation~(\ref{eq:Decay Muster}) to the numerical values of $\langle\hat{n}_{0}\hat{n}_{r}\rangle$ as well as $\langle\hat{c}_{0}^{\dagger}\hat{c}_{r}\rangle$, both calculated with an MPS bond dimension $\chi=256$. The fits were performed in the interval $r=10\dots 60$ with only two free parameters, namely ${\rm const}_{1/2}$ and $K$ of equation~(\ref{eq:Decay Muster}).\label{fig:Luttinger K}}
\end{figure}

\begin{figure}
\includegraphics[width=1\columnwidth]{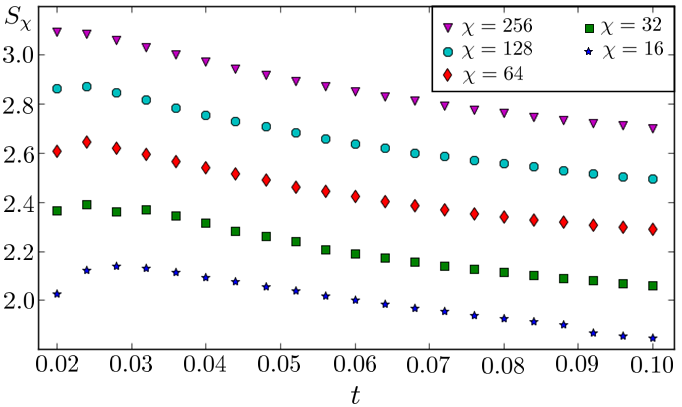}
\caption{The entanglement entropy of the half chain $S_\chi$ extracted from MPS with different bond dimension $\chi$ is plotted in
dependence of the hopping amplitude $t$ for $U=V=1$ and $\mu=1.55$. The different curves are roughly equidistant,
as predicted by equation~(\ref{eq:Entropy Scaling}).
\label{fig:Entropien}}
\end{figure}

\subsubsection{Supersolids for $U=V=1$\label{sub:Supersolid}}
A supersolid is characterized as spatially ordered phase which also
exhibits superfluid properties. We already mentioned the spatial order,
belonging to Figure~\ref{fig:X0Phase}, which is given by the occupation pattern
$\dots,x_{i},\varepsilon_{i+1},x_{i+2},\varepsilon_{i+3},\dots$.
In our numerical studies we consider translational invariant superpositions
of the ground states, where the occupation pattern is still visible
in the two-point correlation functions as $\langle\hat{n}_{0}\hat{n}_{r}\rangle$
and $\langle\hat{c}_{0}^{\dagger}\hat{c}_{r}\rangle$ shown in Figure~\ref{fig:NN_Korrelation}
and Figure~\ref{fig:CC_Korrelation}. Due to the spacial order both correlation functions are
split into two branches, where correlations belonging to odd distances
$r=2j+1$ are strongly suppressed compared to correlations belonging to
even distances $r=2j$. Furthermore, both branches exhibit an algebraic decay
of the same power. For a superfluide, the power of the decay of $\langle\hat{n}_{0}\hat{n}_{r}\rangle$
is supposed to be reciprocal to the power of $\langle\hat{c}_{0}^{\dagger}\hat{c}_{r}\rangle$.
Luttinger liquid theory predicts \cite{Giamarchi2003}
\begin{eqnarray}
\langle\hat{n}_{0}\hat{n}_{r}\rangle & = & \langle\hat{n}_{0}\hat{n}_{r\rightarrow\infty}\rangle+{\rm const}_{1}\cdot\cos\left(2\pi\rho_{0}r\right)\cdot r^{-2K}+\dots\nonumber \\
\langle\hat{c}_{0}^{\dagger}\hat{c}_{r}\rangle & = & {\rm const}_{2}\cdot r^{-\frac{1}{2K}}+\dots\label{eq:Decay Muster}
\end{eqnarray}
with $\rho_{0}=\langle\hat{n}_{i}\rangle$. These relations are confirmed by our numerical results, as displayed in Figure~\ref{fig:Luttinger K}. Both correlation functions give rise to the same Luttinger liquid parameter $K$ within a small error range.

In the thermodynamical limit algebraically decaying correlation functions
go hand in hand with an infinite entanglement entropy of the half chain
$S$, which can not be represented by any MPS with finite bond dimension
$\chi$. Nonetheless, it was shown \cite{Tagliacozzo2008, Pollmann2009} that the numerically obtained
entropy $S_\chi$ for such critical phases shows a predictable scaling
as function of the MPS bond dimension $\chi$ \cite{Pollmann2009}
\begin{equation}
S_{\chi_{2}}-S_{\chi_{1}}\approx\left(\sqrt{\frac{12}{c}}+1\right)^{-1}\cdot\log_{2}\left(\frac{\chi_{2}}{\chi_{1}}\right)\label{eq:Entropy Scaling}
\end{equation}
where $c$ represents the central charge. A demonstration of the scaling
behavior is given in Figure~\ref{fig:Entropien}. From this numerical sample one obtains $S_{\chi=256}-S_{\chi=16}=(0.218\pm 0.003)\times 4$,
which matches  $S_{\chi=256}-S_{\chi=16}\approx0.224\times 4$ drawn from equation~(\ref{eq:Entropy Scaling}) for $c=1$.

\section{Conclusion\label{sec:Conclusion}}

We have presented several extensions to the basic iMPS algorithm for
systems with long-range interactions, some of them with the potential
to be useful in a much broader context. A special focus was set
on problems arising from broken translational invariance. Here, convergence
was ensured by various means, but mainly due to the SMO method which
irons out local variations by optimizing exponentially many MPS simultaneously.
The new algorithm was successfully applied to calculate detailed Devil's
Staircases and phase diagrams for polar bosons \eqref{eq:Bose-Hubbard Ham}
and was also suitable to verify supersolid properties.
Theoretical restraints as considered in the comments \ref{sub:Comments}
seem to have negligible influence on practical applications such that
the new version of the iMPS algorithm is a genuine improvement in
the sense that it can all the old version could plus more.

\section*{Acknowledgments}

This research was funded by the Austrian Science Fund (FWF): P20748-N16, P24273-N16, 
SFB F40-FoQus F4012-N16, the European Union (NAMEQUAM) and by the
Austrian Ministry of Science BMWF as part of the UniInfrastrukturprogramm
of the Research Platform Scientific Computing at the University of
Innsbruck. We further like to thank Andrew J. Daley, Marcello Dalmonte,
Johannes Schachenmayer, Ulrich Schollw\"ock and especially Andreas L\"auchli
for helpful discussions.

\appendix

\section{MPO representation for Hamiltonians\label{sub:Appendix:-MPO-representation}}

For self-consistency we give a short account based on some examples
how to construct an MPO representation for a given Hamiltonian (see
also \cite{Crosswhite2008_Automata,Froewis2010,PolyExp,McCulloch2007_HamMPO}).

For finite systems with open boundaries the Hamiltonian can be written
\eqref{eq:MPO}
\begin{eqnarray}
\mathcal{H}_{s_{1}s_{2}\cdots s_{n}}^{s'_{1}s'_{2}\cdots s'_{n}} & = & H_{s'_{1}s_{1}}^{[1]\,\mu_{1}}\cdot H_{s'_{2}s_{2}}^{[2]\,\mu_{1}\mu_{2}}\cdot H_{s'_{3}s_{3}}^{[3]\,\mu_{2}\mu_{3}}\dots\label{eq:AppendixMPODef}\\
 &  & \cdots H_{s'_{n-1}s_{n-1}}^{[n-1]\,\mu_{n-2}\mu_{n-1}}\cdot H_{s'_{n}s_{n}}^{[n]\,\mu_{n-1}}.\nonumber
\end{eqnarray}
First, we need a neat way to write down the explicit form of the fourth
order tensors $H_{s's}^{\mu_{l}\mu_{r}}$. We write them as matrices
whose entries are matrices, too
\begin{equation}
H_{s's}^{\mu_{l}\mu_{r}}=\left(H^{s's}\right)^{\mu_{l}\mu_{r}}.
\end{equation}
As an example we consider the Ising Hamiltonian
\begin{equation}
\mathcal{H}=-\sum_{i=1}^{n-1}\sigma_{z}^{[i]}\cdot\sigma_{z}^{[i+1]}-\sum_{i=1}^{n}\sigma_{x}^{[i]}.
\end{equation}
As we will see below, a possible choice for all $H^{[k]}$ in equation
\eqref{eq:AppendixMPODef} with $2\leq k\leq n-1$ is
\begin{equation}
\left(H^{s's}\right)^{\mu_{l}\mu_{r}}=\left(\begin{array}{ccc}
\mathbb{I}^{s's} & 0^{s's} & 0^{s's}\\
-\sigma_{z}^{s's} & 0^{s's} & 0^{s's}\\
-\sigma_{x}^{s's} & \sigma_{z}^{s's} & \mathbb{I}^{s's}
\end{array}\right).\label{eq:HMitte}
\end{equation}
$H_{s'_{1}s_{1}}^{[1]\,\mu_{1}}$ and $H_{s'_{n}s_{n}}^{[n]\,\mu_{n-1}}$
are vectors over matrices
\begin{equation}
H_{s'_{1}s_{1}}^{[1]\,\mu_{1}}=\left(\begin{array}{ccc}
-\sigma_{x}^{s_{1}'s_{1}}, & \sigma_{z}^{s_{1}'s_{1}}, & \mathbb{I}^{s_{1}'s_{1}}\end{array}\right).
\end{equation}
In order to get a better understanding we look at the tensor product
of the first $k$ tensors. Below \eqref{eq:Induktion} we show by
induction that
\begin{align}
H^{[1\dots k]} & =H_{s'_{1}s_{1}}^{[1]\,\mu_{1}}\cdot H_{s'_{2}s_{2}}^{[2]\,\mu_{1}\mu_{2}}\cdots H_{s'_{k}s_{k}}^{[k]\,\mu_{k-1}\mu_{k}}\nonumber \\
 & =\left(\begin{array}{ccc}
\left[-\sum_{i=1}^{k-1}\sigma_{z}^{[i]}\cdot\sigma_{z}^{[i+1]}-\sum_{i=1}^{k}\sigma_{x}^{[i]}\right], & \sigma_{z}^{[k]}, & \mathbb{I}\end{array}\right)\nonumber \\
 & =\left(\begin{array}{ccc}
\mathcal{H}^{[k]}, & \sigma_{z}^{[k]}, & \mathbb{I}\end{array}\right)\label{eq:TransportVektor}
\end{align}
The resulting vector $H^{[1\dots k]}$ can be seen as an object with
three ``slots'' in which all the relevant information about the
first $k$ sites are stored. Of course, the number of slots corresponds
to the bond dimension of the MPO. The first slot contains all interaction
terms between the first $k$ sites only and local terms. Since the
$k$th site also interacts with the $k+1$th site, the second slot
of the vector passes on $\sigma_{z}^{[k]}$ and finally, the third
slot preserves the identity $\mathbb{I}=\mathbb{I}^{s_{1}'s_{1}}\otimes\mathbb{I}^{s_{2}'s_{2}}\otimes\cdots\otimes\mathbb{I}^{s_{k}'s_{k}}$.
For $H^{[1]}$ this description is easily checked. The tensor $H^{[k]}=\left(H^{s's}\right)^{\mu_{l}\mu_{r}}$
\eqref{eq:HMitte} is designed such that it performs the correct induction
step

\begin{align}
H^{[1\dots k]} & =H^{[1\dots k-1]}\cdot H^{[k]}\nonumber \\
 & =\left(\begin{array}{ccc}
\mathcal{H}^{[k-1]}, & \sigma_{z}^{[k-1]}, & \mathbb{I}\end{array}\right)\left(\begin{array}{ccc}
\mathbb{I}^{s_{k}'s_{k}} & 0^{s_{k}'s_{k}} & 0^{s_{k}'s_{k}}\\
-\sigma_{z}^{s_{k}'s_{k}} & 0^{s_{k}'s_{k}} & 0^{s_{k}'s_{k}}\\
-\sigma_{x}^{s_{k}'s_{k}} & \sigma_{z}^{s_{k}'s_{k}} & \mathbb{I}^{s_{k}'s_{k}}
\end{array}\right)\nonumber \\
 & =\left(\begin{array}{ccc}
\left[\mathcal{H}^{[k-1]}-\sigma_{z}^{[k-1]}\cdot\sigma_{z}^{[k]}-\sigma_{x}^{[k]}\right], & \sigma_{z}^{[k]}, & \mathbb{I}\end{array}\right)\nonumber \\
 & =\left(\begin{array}{ccc}
\mathcal{H}^{[k]}, & \sigma_{z}^{[k]}, & \mathbb{I}\end{array}\right)\label{eq:Induktion}
\end{align}

The final tensor $H_{s'_{n}s_{n}}^{[n]\,\mu_{n-1}}$is given by
\begin{equation}
H_{s'_{n}s_{n}}^{[n]\,\mu_{n-1}}=\left(\begin{array}{ccc}
\mathbb{I}^{s_{n}'s_{n}}, & -\sigma_{z}^{s_{n}'s_{n}},- & \sigma_{x}^{s_{n}'s_{n}}\end{array}\right)^{T}.
\end{equation}
With that we get
\begin{eqnarray}
H^{[1\dots n]} & = & H^{[1\dots n-1]}\cdot H^{[n]}\nonumber \\
 & = & \left(\begin{array}{ccc}
\mathcal{H}^{[n-1]}, & \sigma_{z}^{[n-1]}, & \mathbb{I}\end{array}\right)\left(\begin{array}{c}
\mathbb{I}^{s_{n}'s_{n}}\\
-\sigma_{z}^{s_{n}'s_{n}}\\
-\sigma_{x}^{s_{n}'s_{n}}
\end{array}\right)\nonumber \\
 & = & \mathcal{H}^{[n-1]}-\sigma_{z}^{[n-1]}\cdot\sigma_{z}^{[n]}-\sigma_{x}^{[n]}\nonumber \\
 & =- & \sum_{i=1}^{n-1}\sigma_{z}^{[i]}\cdot\sigma_{z}^{[i+1]}-\sum_{i=1}^{n}\sigma_{x}^{[i]}.\label{eq:EndHam}
\end{eqnarray}

We described the vector $H^{[1\dots k]}$ \eqref{eq:TransportVektor}
as an object which contains all relevant information of the sites
$1\dots k$. This description is true not only for Ising interaction.
For any Hamiltonian we have to identify what these relevant information
are and design the vector accordingly. As convention, we use the first
slot of the vector to store $\mathcal{H}^{[k]}$ the sum of all interaction
terms between the first $k$ sites only, including local terms. In
the last slot we pass on the identity. The slots in between are needed
for interaction terms which involve (at least) one of the first $k$
sites and (at least) one of the other sites $k+1\dots n$. In the
case of an Heisenberg chain
\begin{equation}
\mathcal{H}=\sum_{i=1}^{n-1}J_{x}\cdot\sigma_{x}^{[i]}\cdot\sigma_{x}^{[i+1]}+\sum_{i=1}^{n-1}J_{y}\cdot\sigma_{y}^{[i]}\cdot\sigma_{y}^{[i+1]}+\sum_{i=1}^{n-1}J_{z}\cdot\sigma_{z}^{[i]}\cdot\sigma_{z}^{[i+1]}
\end{equation}
the vector $H^{[1\dots k]}$ needs five slots
\begin{equation}
H^{[1\dots k]}=\left(\begin{array}{ccccc}
\mathcal{H}^{[k]}, & \sigma_{x}^{[k]}, & \sigma_{y}^{[k]}, & \sigma_{z}^{[k]}, & \mathbb{I}\end{array}\right).
\end{equation}
Further slots might be necessary if we do not restrict ourselves to
nearest neighbor interactions.

Once we have identified the structure of $H^{[1\dots k]}$, it is straight
forward to write the first tensor $H^{[1]}$ in vector form.
The matrix structure of all the following tensors $H^{[j]}$ is constructed
column-wise such that the induction
\begin{equation}
H^{[1\dots j]}=H^{[1\dots j-1]}\cdot H^{[j]}
\end{equation}
is accomplished as in equation \eqref{eq:Induktion} or equation \eqref{eq:EndHam}
for the final tensor $H^{[n]}$. According to our convention, local
terms, as needed in \ref{sub:Energy-overgrow}, are always
represented in the bottom left entry of the matrices.

For long-range interaction, the recipe given so far becomes problematic.
The longer the range of the interaction, the more information has
to be stored in the vector $H^{[1\dots k]}$, which usually requires
more and more slots. But there are some exceptions (see e.g. \cite{Froewis2010}).
An exponentially decaying interaction needs only one slot \--- even
for infinite range. As an example, we look at the toy Hamiltonian

\begin{equation}
\mathcal{H}=J\cdot\sum_{i=1}^{n}\sum_{j=1}^{i-1}\lambda^{i-j-1}\cdot\sigma_{z}^{[j]}\cdot\sigma_{z}^{[i]},
\end{equation}
where $i-j-1$ is the exponent of $\lambda$ and not an index. First,
we have to identify the structure of $H^{[1\dots k]}$
\[
H^{[1\dots k]}=\left(\begin{array}{ccc}
\mathcal{H}^{[k]}, & \sum_{j=1}^{k}\lambda^{k-j}\cdot\sigma_{z}^{[j]}, & \mathbb{I}\end{array}\right).
\]
The crucial observation is, that $\sum_{j=1}^{k}\lambda^{k-j}\cdot\sigma_{z}^{[j]}$
can be generated iteratively. The following tensors fulfill this task
\begin{eqnarray}
H^{[1]} & = & \left(\begin{array}{ccc}
0^{s_{1}'s_{1}}, & \sigma_{z}^{s_{1}'s_{1}}, & \mathbb{I}^{s_{1}'s_{1}}\end{array}\right)\\
H^{[k]} & = & \left(\begin{array}{ccc}
\mathbb{I}^{s_{k}'s_{k}} & 0^{s_{k}'s_{k}} & 0^{s_{k}'s_{k}}\\
J\cdot\sigma_{z}^{s_{k}'s_{k}} & \lambda\cdot\mathbb{I}^{s_{k}'s_{k}} & 0^{s_{k}'s_{k}}\\
0^{s_{k}'s_{k}} & \sigma_{z}^{s_{k}'s_{k}} & \mathbb{I}^{s_{k}'s_{k}}
\end{array}\right)\\
H^{[n]} & = & \left(\begin{array}{ccc}
\mathbb{I}^{s_{n}'s_{n}}, & J\cdot\sigma_{z}^{s_{n}'s_{n}}, & 0^{s_{n}'s_{n}}\end{array}\right)^{T}.
\end{eqnarray}

In order to encode the polynomial decay as $(r)^{-3}$ into an MPO
we resorted to an approximation as a weighted sum of $N_{\exp}$ different
exponential terms, i.e.\
\begin{equation}
(r)^{-3}\approx\sum_{i=1}^{N_{\exp}}a_{i}\cdot\lambda_{i}^{r-1}\quad r=1,2,3,...\label{eq:r-cube}
\end{equation}
In the appendix of \cite{PolyExp} it is shown how to calculate the
optimal $a_{i}$ and $\lambda_{i}$. For the quality of the approximation
in dependence of $N_{\exp}$ see \cite{Crosswhite2008}.

\section{Gain function\label{sub:Appendix:-Gain-function}}

We like to estimate the influence of $\gamma$ in
\begin{equation}
\mathbb{\widetilde{H}}^{[\gamma]}=\mathbb{\widetilde{H}}-\gamma\cdot|A^{[\textrm{refer}]}\rangle\langle A^{[\textrm{refer}]}|\quad\textrm{with}\quad\gamma\geq0.\label{eq:Appendix H gamma}
\end{equation}
on $A^{[\gamma]}$ which is supposed to minimize $\langle A^{[\gamma]}|\mathbb{\widetilde{H}}^{[\gamma]}|A^{[\gamma]}\rangle$.
For our numerical purpose the following simple approximation suffices
\begin{equation}
A^{[\gamma]}=\sqrt{1-\varepsilon^{2}(\gamma)}\cdot A^{[\textrm{refer}]}+\varepsilon(\gamma)\cdot B,\label{eq:Appendix A Gamma}
\end{equation}
with $\Vert A^{[\textrm{refer}]}\Vert$=$\Vert B\Vert=1$ and $A^{[\textrm{refer}]}\perp B=\textrm{const}$.
The vector $B$ is extracted from $A^{[\gamma=0]}$, which has to
be calculated first. This might seem inefficient, since we have to
minimize $\langle A|\mathbb{\widetilde{H}}^{[\gamma]}|A\rangle$ twice
\--- once for $\gamma=0$ and once for the final value of $\gamma$.
The solution is to use a minimization routine which projects the minimization
onto a small subspace, as explained in section \ref{sub:Minimization-routine-and}.
This projection has to be done only once but can be used twice. Since
the projection is the most time consuming part, the double calculation
is done quite cheap. More to this at the end of this subsection.

Now, we calculate the pseudo energy $E$ using the equations \eqref{eq:Appendix H gamma}
and \eqref{eq:Appendix A Gamma}.

\begin{eqnarray}
E & = & \langle A^{[\gamma]}|\mathbb{\widetilde{H}}^{[\gamma]}|A^{[\gamma]}\rangle\nonumber \\
 & = & \left(1-\varepsilon^{2}\right)\cdot\left(\langle A^{[\textrm{refer}]}|\mathbb{\widetilde{H}}|A^{[\textrm{refer}]}\rangle-\gamma\right)+\varepsilon^{2}\cdot\langle B|\mathbb{\widetilde{H}}|B\rangle\nonumber \\
 & + & 2\cdot\varepsilon\cdot\sqrt{1-\varepsilon\text{\texttwosuperior}}\textrm{ Re}\langle A^{[\textrm{refer}]}|\mathbb{\widetilde{H}}|B\rangle.
\end{eqnarray}
In the following we approximate $\varepsilon\cdot\sqrt{1-\varepsilon\text{\texttwosuperior}}\thickapprox\varepsilon$.
This approximation is not needed but it keeps the calculations clear.
In addition, the formula we will derive from the approximated version
is numerically more stable. In our program we used the exact version
(which we will not derive here) only if $\varepsilon(\gamma=0)>0.01$.

With $\varepsilon\cdot\sqrt{1-\varepsilon\text{\texttwosuperior}}\thickapprox\varepsilon$,
$E$ is a parabola in $\varepsilon$. Assuming that the apex is the
minimum one gets

\begin{equation}
\varepsilon_{\textrm{min}}(\gamma)=\frac{-\textrm{ Re}\langle A^{[\textrm{refer}]}|\mathbb{\widetilde{H}}|B\rangle}{\langle B|\mathbb{\widetilde{H}}|B\rangle-\langle A^{[\textrm{refer}]}|\mathbb{\widetilde{H}}|A^{[\textrm{refer}]}\rangle+\gamma}.
\end{equation}
Since $\Vert A_{[n]}^{[\gamma]}-A_{[n]}^{[\textrm{refer}]}\Vert\approx\varepsilon_{\textrm{min}}$
for $\varepsilon_{\textrm{min}}\ll1$, we choose $\varepsilon_{\textrm{min}}$
in accordance with equation \eqref{eq:C_n und D_max} to be

\begin{equation}
\varepsilon_{\textrm{min}}(\gamma)=\min\left(c\cdot\varepsilon_{\textrm{min}}(\gamma=0),\Delta_{\textrm{max}}\right).
\end{equation}
From that $\gamma$ is calculated to be
\begin{eqnarray}
\gamma & = & \max\left(\gamma_{[c]},\gamma_{[\Delta_{\textrm{max}}]}\right)\quad\textrm{with}\nonumber \\
\gamma_{[c]} & = & \frac{1-c}{c}\cdot\left(\langle B|\mathbb{\widetilde{H}}|B\rangle-\langle A^{[\textrm{refer}]}|\mathbb{\widetilde{H}}|A^{[\textrm{refer}]}\rangle\right)\nonumber \\
\gamma_{[\Delta_{\textrm{max}}]} & = & \langle A^{[\textrm{refer}]}|\mathbb{\widetilde{H}}|A^{[\textrm{refer}]}\rangle-\langle B|\mathbb{\widetilde{H}}|B\rangle\nonumber \\
 & - & \frac{1}{\Delta_{\textrm{max}}}\cdot\textrm{ Re}\langle A^{[\textrm{refer}]}|\mathbb{\widetilde{H}}|B\rangle.
\end{eqnarray}

\subsection{Subspace projection and $\gamma$\label{sub:Subspace-projection-and}}

As mentioned, we have to solve $\min\langle A^{[\gamma]}|\mathbb{\widetilde{H}}^{[\gamma]}|A^{[\gamma]}\rangle$
twice: first for $\gamma=0$ and after that for the final value of
$\gamma$. The idea is to reuse the information gathered in the first
minimization for the second run. As described in \ref{sub:Minimization-routine-and},
the minimization problem is projected onto a subspace. The first basis
vector of this subspace is $|\mathfrak{A}_{1}\rangle=|A^{[\textrm{refer}]}\rangle$
\eqref{eq:First A}. Hence the only element of the subspace matrix
$\mathfrak{H}$ \eqref{eq:Subspace-Matrix} which has to be adopted
is $\mathfrak{H}_{1,1}\leftarrow\mathfrak{H}_{1,1}-\gamma$.

So, the update of the subspace matrix $\mathfrak{H}$ is extremely
simple to perform but one might wonder whether the associated basis
vectors $|\mathfrak{A}_{i}\rangle$ are still optimal. For the Lanczos
\cite{Lanczos1951} and Arnoldi \cite{Arnoldi1951} algorithms, which
are built on pure Krylov spaces, it turns out that the influence of
$\gamma$ on the basis vectors gets extinguished, while for the extended
algorithm presented in \ref{sub:Minimization-routine-and} the optimal
choice of the basis vectors shows a slight dependence on $\gamma$.
Numerically, this is not a serious problem. Nonetheless, since the
second optimization is the important one we recommend to use an approximated
value of the final $\gamma$ to construct the basis vectors in the
first run.  A simple and effective way is to use $\gamma_{[n-1]}$
from the last tensor optimization assuming $\gamma_{[n]}\approx\gamma_{[n-1]}$.
Alternatively, one can solve the intermediate subspace matrices $\mathfrak{H}$
and use these intermediate results to calculate approximations of
$\gamma$ as described above.

\section{Transformation proof for degenerate ground states\label{sub:Appendix:-Transformation-poof}}

Let $\mathcal{A}$ and $\mathcal{B}$ represent two different ground
states to the same Hamiltonian with a $g$ times degenerate ground
state level due to broken translational invariance
\begin{eqnarray}
\mathcal{A} & = & \cdots Q_{[L]\, s_{-1}}^{\alpha_{-2}\alpha_{-1}}\cdot Q_{[L]\, s_{0}}^{\alpha_{-1}\widetilde{\alpha}_{0}}\cdot\lambda^{\tilde{\alpha}_{0}\alpha_{0}}\cdot Q_{[R]\, s_{1}}^{\alpha_{0}\alpha_{1}}\cdot Q_{[R]\, s_{2}}^{\alpha_{1}\alpha_{2}}\cdots\nonumber \\
\mathcal{B} & = & \cdots q_{[L]\, s_{-1}}^{\alpha_{-2}\alpha_{-1}}\cdot q_{[L]\, s_{0}}^{\alpha_{-1}\widetilde{\alpha}_{0}}\cdot\xi^{\tilde{\alpha}_{0}\alpha_{0}}\cdot q_{[R]\, s_{1}}^{\alpha_{0}\alpha_{1}}\cdot q_{[R]\, s_{2}}^{\alpha_{1}\alpha_{2}}\cdots\label{eq:Teo Ground states}
\end{eqnarray}
If degenerations due to further symmetries are involved, $\mathcal{A}$
and $\mathcal{B}$ are supposed to have the same characteristic values
for these symmetries. This allows us to operate as if no further symmetries
exist, since all operations we are about to use leave these characteristic
values unchanged. Under this condition we will prove the existence
of a matrix $\gamma^{\tilde{\alpha}_{0}\alpha_{0}}$ such that the
ground state $\mathcal{B}$ can be expressed using the tensors $Q$
stemming from the iMPS $\mathcal{A}$
\begin{equation}
\mathcal{B}=\cdots Q_{[L]\, s_{-1}}^{\alpha_{-2}\alpha_{-1}}\cdot Q_{[L]\, s_{0}}^{\alpha_{-1}\widetilde{\alpha}_{0}}\cdot\gamma^{\tilde{\alpha}_{0}\alpha_{0}}\cdot Q_{[R]\, s_{1}}^{\alpha_{0}\alpha_{1}}\cdot Q_{[R]\, s_{2}}^{\alpha_{1}\alpha_{2}}\cdots\label{eq:Alternative Zustand Form}
\end{equation}
where $\gamma^{\tilde{\alpha}_{0}\alpha_{0}}=\omega^{\tilde{\alpha}_{0}\beta}\cdot\xi^{\beta\delta}\cdot\nu^{\delta\alpha_{0}}$
.

Let us start by surveying the elements of the proof. In order to show
the claimed equation \eqref{eq:Alternative Zustand Form} we will
actually prove the gauge transformation
\begin{eqnarray}
\cdots q_{[L]\, s_{-1}}^{\alpha_{-2}\alpha_{-1}}\cdot q_{[L]\, s_{0}}^{\alpha_{-1}\widetilde{\alpha}_{0}} & = & \cdots Q_{[L]\, s_{-1}}^{\alpha_{-2}\alpha_{-1}}\cdot Q_{[L]\, s_{0}}^{\alpha_{-1}\beta}\cdot\omega^{\beta\widetilde{\alpha}_{0}}\nonumber \\
q_{[R]\, s_{1}}^{\alpha_{0}\alpha_{1}}\cdot q_{[R]\, s_{2}}^{\alpha_{1}\alpha_{2}}\cdots & = & \nu^{\alpha_{0}\beta}\cdot Q_{[R]\, s_{1}}^{\beta\alpha_{1}}\cdot Q_{[R]\, s_{2}}^{\alpha_{1}\alpha_{2}}\cdots\label{eq:EichGleichung}
\end{eqnarray}
This gauge transformation will be proven for the case that the two
underlying MPS represent the same physical state \--- which is not
given for $\mathcal{A}$ and $\mathcal{B}$ \eqref{eq:Teo Ground states}.
In order to use the gauge proof for our purpose we need to find one
physical state described by two different MPS where the first MPS
be constructed using the $Q$ of $\mathcal{A}$ and the second by
the $q$ of $\mathcal{B}$. This one state which allows us to complete
our proof is the translational invariant ground state. According to
our preliminary remarks this state is unique. Hence, if we succeed
to construct two different MPS which represent a translational invariant
ground state, we know that they represent the same physical state
as demanded. We will prove the following construction for this state

\begin{eqnarray}
\mathcal{T}_{Q} & = & \cdots Q_{[L]\, s_{-1}}^{\alpha_{-2}\alpha_{-1}}\cdot Q_{[L]\, s_{0}}^{\alpha_{-1}\widetilde{\alpha}_{0}}\cdot\tau^{\tilde{\alpha}_{0}\alpha_{0}}\cdot Q_{[R]\, s_{1}}^{\alpha_{0}\alpha_{1}}\cdot Q_{[R]\, s_{2}}^{\alpha_{1}\alpha_{2}}\cdots\nonumber \\
 & = & \cdots q_{[L]\, s_{-1}}^{\alpha_{-2}\alpha_{-1}}\cdot q_{[L]\, s_{0}}^{\alpha_{-1}\widetilde{\alpha}_{0}}\cdot\theta^{\tilde{\alpha}_{0}\alpha_{0}}\cdot q_{[R]\, s_{1}}^{\alpha_{0}\alpha_{1}}\cdot q_{[R]\, s_{2}}^{\alpha_{1}\alpha_{2}}\cdots\nonumber \\
 & = & \mathcal{T}_{q}\label{eq:Gleicher Trans Inv Grundstate}
\end{eqnarray}
with new tensors $\tau{}^{\tilde{\alpha}_{0}\alpha_{0}}$ and $\theta^{\tilde{\alpha}_{0}\alpha_{0}}$.

We start by showing the gauge transformation \eqref{eq:EichGleichung}.
In order to increase clarity we define
\begin{eqnarray}
\overrightarrow{\mathbb{Q}}{}_{s_{1}\cdots s_{k}}^{\alpha_{k}} & = & Q_{[1]\, s_{1}}^{\alpha_{1}}\cdots Q_{[n]\, s_{k}}^{\alpha_{k-1}\alpha_{k}}\nonumber \\
\overrightarrow{\mathbb{P}}{}_{s_{1}\cdots s_{k}}^{\alpha_{k}} & = & p_{[1]\, s_{1}}^{\alpha_{1}}\cdots p_{[n]\, s_{k}}^{\alpha_{k-1}\alpha_{k}}.\label{eq:Q und q Pfeil}
\end{eqnarray}
For this specific gauge proof the different $Q_{[i]}$ do not need
to be of the same structure and neither do the $p_{[i]}$. But in
contrast to the $p_{[i]}$ the $Q_{[i]}$ still have to fulfill equation
\eqref{eq:OrthogonalQ}, i.e.,
\begin{eqnarray}
\overrightarrow{\mathbb{Q}}{}_{s_{1}\cdots s_{k}}^{*\beta_{k}}\cdot\overrightarrow{\mathbb{Q}}{}_{s_{1}\cdots s_{k}}^{\alpha_{k}} & = & \delta{}^{\beta_{k}\alpha_{k}},\:\textrm{while}\label{eq:QQ-Identitat}\\
\overrightarrow{\mathbb{Q}}{}_{\bar{s}_{1}\cdots\bar{s}_{k}}^{\alpha_{k}}\cdot\overrightarrow{\mathbb{Q}}{}_{s_{1}\cdots s_{k}}^{*\alpha_{k}} & \neq & \delta_{\left(\bar{s}_{1}\cdots\bar{s}_{k}\right)\left(s_{1}\cdots s_{k}\right)},\nonumber
\end{eqnarray}
Although not an identity, the operator $\overrightarrow{\mathbb{Q}}{}_{\bar{s}_{1}\cdots\bar{s}_{k}}^{\alpha_{k}}\cdot\overrightarrow{\mathbb{Q}}{}_{s_{1}\cdots s_{k}}^{*\alpha_{k}}$
acts like such if it is applied from the left on any MPS with the
structure $\overrightarrow{\mathbb{Q}}{}_{s_{1}\cdots s_{k}}^{\alpha_{k}}\cdot\overleftarrow{\mathbb{R}}{}_{s_{k+1}\dots s_{n}}^{\alpha_{k}}$,
where $\overleftarrow{\mathbb{R}}{}_{s_{k+1}\dots s_{n}}^{\alpha_{k}}$
represents an arbitrary right side of the MPS. Using the identity
\eqref{eq:QQ-Identitat} we get
\begin{alignat}{1}
 & \left(\overrightarrow{\mathbb{Q}}{}_{\bar{s}_{1}\cdots\bar{s}_{k}}^{\beta_{k}}\cdot\overrightarrow{\mathbb{Q}}{}_{s_{1}\cdots s_{k}}^{*\beta_{k}}\right)\cdot\left(\overrightarrow{\mathbb{Q}}{}_{s_{1}\cdots s_{k}}^{\alpha_{k}}\cdot\overleftarrow{\mathbb{R}}{}_{s_{k+1}\dots s_{n}}^{\alpha_{k}}\right)\nonumber \\
 & =\overrightarrow{\mathbb{Q}}{}_{\bar{s}_{1}\cdots\bar{s}_{k}}^{\beta_{k}}\cdot\delta^{\beta_{k}\alpha_{k}}\cdot\overleftarrow{\mathbb{R}}{}_{s_{k+1}\dots s_{n}}^{\alpha_{k}}\nonumber \\
 & =\overrightarrow{\mathbb{Q}}{}_{s_{1}\cdots s_{k}}^{\alpha_{k}}\cdot\overleftarrow{\mathbb{R}}{}_{s_{k+1}\dots s_{n}}^{\alpha_{k}}\label{eq:Pseudo Identity}
\end{alignat}
In the following we assume that the two MPS
\begin{equation}
\overrightarrow{\mathbb{Q}}{}_{s_{1}\cdots s_{k}}^{\alpha_{k}}\cdot\overleftarrow{\mathbb{R}}{}_{s_{k+1}\dots s_{n}}^{\alpha_{k}}=\overrightarrow{\mathbb{P}}{}_{s_{1}\cdots s_{k}}^{\alpha_{k}}\cdot\overleftarrow{\mathbb{M}}{}_{s_{k+1}\dots s_{n}}^{\alpha_{k}}\label{eq:Gleicher Phys Zustand}
\end{equation}
 represent the same physical state. Now, let us apply the operator
$\overrightarrow{\mathbb{Q}}{}_{\bar{s}_{1}\cdots\bar{s}_{k}}^{\beta_{k}}\cdot\overrightarrow{\mathbb{Q}}{}_{s_{1}\cdots s_{k}}^{*\beta_{k}}$
on this equation. Due to equation \eqref{eq:Pseudo Identity} the
left side stays unchanged. Hence the physical state represented by
the right side does not change either

\begin{alignat}{1}
 & \overrightarrow{\mathbb{P}}{}_{s_{1}\cdots s_{k}}^{\alpha_{k}}\cdot\overleftarrow{\mathbb{M}}{}_{s_{k+1}\dots s_{n}}^{\alpha_{k}}\nonumber \\
 & =\overrightarrow{\mathbb{Q}}{}_{\bar{s}_{1}\cdots\bar{s}_{k}}^{\beta_{k}}\cdot\overrightarrow{\mathbb{Q}}{}_{s_{1}\cdots s_{k}}^{*\beta_{k}}\cdot\overrightarrow{\mathbb{P}}{}_{s_{1}\cdots s_{k}}^{\alpha_{k}}\cdot\overleftarrow{\mathbb{M}}{}_{s_{k+1}\dots s_{n}}^{\alpha_{k}}\nonumber \\
 & =\overrightarrow{\mathbb{Q}}{}_{\bar{s}_{1}\cdots\bar{s}_{k}}^{\beta_{k}}\cdot\omega^{\beta_{k}\alpha_{k}}\cdot\overleftarrow{\mathbb{M}}{}_{s_{k+1}\dots s_{n}}^{\alpha_{k}},\label{eq:Gauge u}
\end{alignat}
with $\omega^{\beta\alpha_{k}}=\overrightarrow{\mathbb{Q}}{}_{s_{1}\cdots s_{k}}^{*\beta_{k}}\cdot\overrightarrow{\mathbb{P}}{}_{s_{1}\cdots s_{k}}^{\alpha_{k}}$.
The existence of the inverse $\overleftarrow{\mathbb{M}}{}^{-1}$
with $\overleftarrow{\mathbb{M}}{}_{s_{k+1}\dots s_{n}}^{\alpha_{k}}\cdot\overleftarrow{\mathbb{M}}{}_{s_{k+1}\dots s_{n}}^{-1\,\beta_{k}}=\delta^{\alpha_{k}\beta_{k}}$
can be assured: if the rank of $\overleftarrow{\mathbb{M}}{}_{(s_{k+1}\dots s_{n})}^{\alpha_{k}}$
is smaller than the bond dimension $\alpha_{k}$, the value of $\alpha_{k}$
can be reduced, since in that case it turns out to be unnecessarily
high. Applying $\overleftarrow{\mathbb{M}}{}^{-1}$ from the right
on equation \eqref{eq:Gauge u} we end up with
\begin{equation}
\overrightarrow{\mathbb{P}}{}_{s_{1}\cdots s_{k}}^{\alpha_{k}}=\overrightarrow{\mathbb{Q}}{}_{s_{1}\cdots s_{k}}^{\beta_{k}}\cdot\omega^{\beta_{k}\alpha_{k}}.\label{eq:Bewiesen Eich U}
\end{equation}
For $k\rightarrow\infty$ this covers the first part of the heralded
gauge equation \eqref{eq:EichGleichung}. The second part of equation
\eqref{eq:EichGleichung} is proved by a straight forward application
of the arguments used above on the right side of the MPS.

Now, we have to show the MPS construction of the translational invariant
ground state \eqref{eq:Gleicher Trans Inv Grundstate}. As above,
we assume that the ground state level of the Hamiltonian under consideration
is $g$ times degenerate due to broken translational invariance.
Let $T_{L}$ be the operator which shifts all sites of an MPS by one
position to the left and $T_{R}=\left(T_{L}\right)^{-1}$. For the
MPS $\mathcal{A}$ representing any of the possible ground states
we get
\begin{eqnarray}
T_{L}\cdot\mathcal{A} & \neq & \mathcal{A}\nonumber \\
\left(T_{L}\right)^{g}\cdot\mathcal{A} & = & \mathcal{A}\nonumber \\
\mathcal{T} & := & \sum_{j=0}^{g-1}\left(T_{L}\right)^{j}\cdot\mathcal{A}\nonumber \\
T_{L}\cdot\mathcal{T} & = & \mathcal{T},\label{eq:Trans inv ueberlagerung}
\end{eqnarray}
where $\mathcal{T}$ represents an unnormalized version of the translational
invariant ground state, which we like to construct. As an intermediate
step we like to prove equation \eqref{eq:Simple Version of Shifted MPS}
below. Therefor we have to look at the effect $T_{L}$ has on $\mathcal{A}$
(using equation \eqref{eq:Teo Ground states} for $\mathcal{A}$ )
\begin{alignat}{1}
 & T_{L}\left(\cdots Q_{[L]\, s_{-1}}^{\alpha_{-2}\alpha_{-1}}\cdot Q_{[L]\, s_{0}}^{\alpha_{-1}\widetilde{\alpha}_{0}}\cdot\mbox{\ensuremath{\lambda^{\tilde{\alpha}_{0}\alpha_{0}}}}\cdot Q_{[R]\, s_{1}}^{\alpha_{0}\alpha_{1}}\cdot Q_{[R]\, s_{2}}^{\alpha_{1}\alpha_{2}}\cdots\right)\nonumber \\
 & =\cdots Q_{[L]\, s_{-1}}^{\alpha_{-2}\tilde{\alpha}_{-1}}\cdot\lambda^{\tilde{\alpha}_{-1}\alpha_{-1}}\cdot Q_{[R]\, s_{0}}^{\alpha_{-1}\alpha_{0}}\cdot Q_{[R]\, s_{1}}^{\alpha_{0}\alpha_{1}}\cdot Q_{[R]\, s_{2}}^{\alpha_{1}\alpha_{2}}\cdots\label{eq:Shifted MPS-1}
\end{alignat}
Next, we look at the MPS $\left(T_{R}\right)^{g}\cdot T_{L}\cdot\mathcal{A}$,
which has the form
\begin{equation}
\cdots Q_{[L]\, s_{-1}}^{\alpha_{-2}\alpha_{-1}}Q_{[L]\, s_{0}}^{\alpha_{-1}\alpha_{0}}\cdots Q_{[L]\, s_{g-1}}^{\alpha_{g-2}\widetilde{\alpha}_{g-1}}\lambda^{\widetilde{\alpha}_{g-1}\alpha_{g-1}}Q_{[R]\, s_{g}}^{\alpha_{g-1}\alpha_{g}}\cdots
\end{equation}
 Since the two MPS $\left(T_{R}\right)^{g}\cdot T_{L}\cdot\mathcal{A}=T_{L}\cdot\mathcal{A}$
describe the same physical state we are allowed to apply the gauge
transformation \eqref{eq:Bewiesen Eich U} and identify
\begin{equation}
\cdots Q_{[L]\, s_{-1}}^{\alpha_{-2}\tilde{\alpha}_{-1}}\cdot\lambda^{\tilde{\alpha}_{-1}\alpha_{-1}}\cdot Q_{[R]\, s_{0}}^{\alpha_{-1}\alpha_{0}}=\cdots Q_{[L]\, s_{-1}}^{\alpha_{-2}\alpha_{-1}}\cdot Q_{[L]\, s_{0}}^{\alpha_{-1}\gamma}\cdot\omega^{\gamma\widetilde{\alpha}_{0}}.\label{eq:Eichung fuer Shift Eins}
\end{equation}
Inserting this expression into equation \eqref{eq:Shifted MPS-1}
we arrive at the following description for $T_{L}\cdot\mathcal{A}$

\begin{alignat}{1}
 & T_{L}\left(\cdots Q_{[L]\, s_{-1}}^{\alpha_{-2}\alpha_{-1}}\cdot Q_{[L]\, s_{0}}^{\alpha_{-1}\widetilde{\alpha}_{0}}\cdot\mbox{\ensuremath{\lambda^{\tilde{\alpha}_{0}\alpha_{0}}}}\cdot Q_{[R]\, s_{1}}^{\alpha_{0}\alpha_{1}}\cdot Q_{[R]\, s_{2}}^{\alpha_{1}\alpha_{2}}\cdots\right)\nonumber \\
 & =\cdots Q_{[L]\, s_{-1}}^{\alpha_{-2}\alpha_{-1}}\cdot Q_{[L]\, s_{0}}^{\alpha_{-1}\widetilde{\alpha}_{0}}\cdot\omega\mbox{\ensuremath{^{\tilde{\alpha}_{0}\alpha_{0}}}}\cdot Q_{[R]\, s_{1}}^{\alpha_{0}\alpha_{1}}\cdot Q_{[R]\, s_{2}}^{\alpha_{1}\alpha_{2}}\cdots\label{eq:Simple Version of Shifted MPS}
\end{alignat}
As we see, applying the operator $T_{L}$ on $\mathcal{A}$ has the
same effect as the replacement of the matrix $\lambda^{\tilde{\alpha}_{0}\alpha_{0}}$
by $\omega\mbox{\ensuremath{^{\tilde{\alpha}_{0}\alpha_{0}}}}$. Since
higher powers of $T_{L}$ can be created by an iteration of the arguments
just presented, it follows that the effect of any $\left(T_{L}\right)^{j}$
on $\mathcal{A}$ can be accounted for by an accordingly calculated
$\omega_{[j]}^{\tilde{\alpha}_{0}\alpha_{0}}$. Following that construction
the only difference between the various MPS $\left(T_{L}\right)^{j}\cdot\mathcal{A}$
is their tensor $\omega_{[j]}^{\tilde{\alpha}_{0}\alpha_{0}}$ and
hence the task of equation \eqref{eq:Trans inv ueberlagerung} to
sum up these MPS reduces to a summation of the $\omega_{[j]}^{\tilde{\alpha}_{0}\alpha_{0}}$.
In other words: Replacing the matrix $\lambda^{\tilde{\alpha}_{0}\alpha_{0}}$
in the MPS $\mathcal{A}$ by $\tau^{\tilde{\alpha}_{0}\alpha_{0}}$
\begin{equation}
\tau^{\tilde{\alpha}_{0}\alpha_{0}}=\sum_{j=0}^{g-1}\omega_{[j]}^{\tilde{\alpha}_{0}\alpha_{0}}\label{eq:Trans inv Rest}
\end{equation}
results in the translational invariant MPS $\mathcal{T}$. Of course,
the same arguments can be applied to the MPS $\mathcal{B}$ giving
us $\mathcal{T}_{Q}=\mathcal{T}=\mathcal{T}_{q}$ as claimed in equation
\eqref{eq:Gleicher Trans Inv Grundstate}.

Let us review our arguments: By virtue of equation \eqref{eq:Trans inv Rest}
we can transform the MPS $\mathcal{A}$ and $\mathcal{B}$ given in
equation \eqref{eq:Teo Ground states} such that we end up with the
translational invariant ground state $\mathcal{T}_{Q}$ and $\mathcal{T}_{q}$
 as claimed in equation \eqref{eq:Gleicher Trans Inv Grundstate}.
Since we work under the condition that $\mathcal{T}_{Q}=\mathcal{T}_{q}$
we are allowed to use the gauge transformation \eqref{eq:Bewiesen Eich U}
to replace the $q$ of $\mathcal{T}_{q}$ by the $Q$ of $\mathcal{T}_{Q}$.
The same replacement is possible in $\mathcal{B}$ because the $q$
in $\mathcal{B}$ and $\mathcal{T}_{q}$ are identical (as are the
$Q$ in $\mathcal{A}$ and $\mathcal{T}_{Q}$). This concludes the
proof of equation \eqref{eq:Alternative Zustand Form} we aimed for.

\section{Davidson implementation\label{sub:Appendix:-Davidson-implementation}}

In this subsection we introduce a practical implementation resembling
the Davidson \cite{DavidsonJacobi} method based on recycled information
of the previous round, which allows to improve the update equation
\eqref{eq:Lanczos update} of the iterative eigenvector solver presented
in section \ref{sub:Minimization-routine-and}. We adopt the same
notation as in that section but mostly drop the index $[n]$ to keep
the formulae clean.

The best possible new vector$|\mathfrak{A}_{k+1}\rangle$ the iterative
eigenvector solver could come up with to replace equation \eqref{eq:Lanczos update}
is an orthonormalized version of $|\Delta_{0}\rangle=|E_{0}\rangle-|e_{0}\rangle$.
\begin{eqnarray}
\mathbb{\widetilde{H}}\cdot|E_{0}\rangle & = & E_{0}\cdot|E_{0}\rangle\nonumber \\
\mathbb{\widetilde{H}}\cdot\left(|e_{0}\rangle+|\Delta_{0}\rangle\right) & = & E_{0}\cdot\left(|e_{0}\rangle+|\Delta_{0}\rangle\right)\nonumber \\
\left(E_{0}\cdot\mathbb{I}-\mathbb{\widetilde{H}}\right)\cdot|\Delta_{0}\rangle & = & \left(\mathbb{\widetilde{H}}-E_{0}\cdot\mathbb{I}\right)\cdot|e_{0}\rangle\nonumber \\
\left(E_{0}\cdot\mathbb{I}-\mathbb{\widetilde{H}}\right)\cdot|\Delta_{0}\rangle & \approx & \left(\mathbb{\widetilde{H}}-e_{0}\cdot\mathbb{I}\right)\cdot|e_{0}\rangle\nonumber \\
\left(E_{0}\cdot\mathbb{I}-\mathbb{\widetilde{H}}\right)\cdot|\Delta_{0}\rangle & = & |r\rangle\nonumber \\
|\Delta_{0}\rangle & = & \left(E_{0}\cdot\mathbb{I}-\mathbb{\widetilde{H}}\right)^{-1}\cdot|r\rangle,\label{eq:DavidsonEinleitung}
\end{eqnarray}
where we used the definition \eqref{eq:residual vector} for $|r\rangle$.
The Davidson method requires a workable approximation for the non
trivial operator $\mathfrak{D}=\left(E_{0}\cdot\mathbb{I}-\mathbb{\widetilde{H}}\right)^{-1}$.
At this point we take advantage of the expectation that the operator
$\mathfrak{D}_{[n]}$ calculated in round $n$ should look pretty
much the same as $\mathfrak{D}_{[n-1]}$ calculated in round $n-1$
\begin{align}
\mathfrak{D}_{[n]} & \approx\mathfrak{D}_{[n-1]}\nonumber \\
\left(E_{[n]\,0}\cdot\mathbb{I}-\mathbb{\widetilde{H}}_{[n]}\right)^{-1} & \approx\left(E_{[n-1]\,0}\cdot\mathbb{I}-\mathbb{\widetilde{H}}_{[n-1]}\right)^{-1}.
\end{align}
Hence we use the accumulated data at the end of round $n-1$ for an
efficient one time estimation of $\mathfrak{D}_{[n-1]}$, which we
will apply in round $n$.

In order to calculate $\mathfrak{D}$ we need a simplified form of
$\mathbb{\widetilde{H}}$ which allows easy inversion. We know $k$
approximated eigenvectors $|e_{i<k}\rangle$ of $\mathbb{\widetilde{H}}$.
In order to have an orthonormal basis for $\mathbb{\widetilde{H}}$
we imagine $N-k$ further $|\hat{e}_{k\leq i<N}\rangle$, where $\dim\left(\mathbb{\widetilde{H}}\right)=N\times N$.
With that we approximate $\mathbb{\widetilde{H}}$ as
\begin{eqnarray}
\mathbb{\widetilde{H}} & \approx & \sum_{i=0}^{k-1}\mathbb{\widetilde{H}}|e_{i}\rangle\langle e_{i}|+\alpha\cdot\sum_{i=k}^{N}|\hat{e}_{i}\rangle\langle\hat{e}_{i}|\nonumber \\
 & = & \sum_{i=0}^{k-1}\left(\mathbb{\widetilde{H}}-\alpha\right)|e_{i}\rangle\langle e_{i}|+\alpha\cdot\mathbb{I}\label{eq:David approx Ham}
\end{eqnarray}
with an average eigenvalue $\alpha=\textrm{const.}$ for the unknown
eigenvectors. One might be tempted to simplify equation \eqref{eq:David approx Ham}
using $\mathbb{\widetilde{H}}|e_{i}\rangle\approx e_{i}\cdot|e_{i}\rangle$
\--- but we do not, since the eigenvectors are not very well approximated
except for $|e_{0}\rangle$. To be able to perform the inversion in
equation \eqref{eq:Davidson Roh-Invers} it suffices to resort to
the exact result

\begin{equation}
\langle e_{i}|\mathbb{\widetilde{H}}|e_{j}\rangle=e_{i}\cdot\delta_{ij}\quad\textrm{without}\sum_{i},\label{eq:David Diagonal}
\end{equation}
which is a consequence of the construction \eqref{eq:Subspace-Matrix}.
Further, with the results gathered during the optimization \eqref{sub:Minimization-routine-and}
the $\mathbb{\widetilde{H}}|e_{i}\rangle$ are as quickly calculated
as the $|e_{i}\rangle$.

Next, we take the trace of equation \eqref{eq:David approx Ham} and
set $\alpha$ such that both sides are equal
\begin{eqnarray}
\textrm{tr}\mathbb{\left(\widetilde{H}\right)} & = & \langle e_{i}|\mathbb{\widetilde{H}}|e_{i}\rangle+\alpha\cdot\left(N-k\right)\nonumber \\
\alpha & = & \frac{\textrm{tr}\mathbb{\left(\widetilde{H}\right)}-\sum_{i=0}^{k-1}e_{i}}{N-k}.
\end{eqnarray}
The trace of the exact $\mathbb{\widetilde{H}}$ is efficiently calculated
by already tracing over its components $L^{\alpha'_{l}\mu_{l}\alpha_{l}}$
and $R^{\alpha'_{r}\mu_{r}\alpha_{r}}$ before assembling them \eqref{eq:Effektive H}.

Now we insert the approximated $\mathbb{\widetilde{H}}$ \eqref{eq:David approx Ham}
in $\mathfrak{D}$
\begin{eqnarray}
\mathfrak{D} & = & \left(E_{0}\cdot\mathbb{I}-\mathbb{\widetilde{H}}\right)^{-1}\nonumber \\
 & \approx & \left(\left(E_{0}-\alpha\right)\cdot\mathbb{I}-\sum_{i=0}^{k-1}\left(\mathbb{\widetilde{H}}-\alpha\right)|e_{i}\rangle\langle e_{i}|\right)^{-1}.\label{eq:Davidson Roh-Invers}
\end{eqnarray}
The inversion is solved by
\begin{equation}
\mathfrak{D}=\left(E_{0}-\alpha\right)^{-1}\cdot\left(\mathbb{I}+\sum_{i=0}^{k-1}\frac{\mathbb{\widetilde{H}}-\alpha}{E_{0}-e_{i}}|e_{i}\rangle\langle e_{i}|\right),\label{eq:Fast-End-Davidson}
\end{equation}
as can be verified inserting the result in $\mathfrak{D}\cdot\mathfrak{D^{-1}}=\mathbb{I}$.

The final question we have to answer is which value we assign to the
unknown exact eigenvalue $E_{0}$. The best approximation (which we
already used in equation \eqref{eq:DavidsonEinleitung}) is $E_{0}\approx e_{0}$,
but this produces a singularity in $\mathfrak{D}$. There are two
ways out. First, we can always pick $E_{0}$ a little bit lower $E_{0}:=e_{0}-\varepsilon$.
Second, we should discard the troublesome term $\sim|e_{0}\rangle\langle e_{0}|$
in $\mathfrak{D}$ anyway for the following reason: We replace equation
\eqref{eq:Lanczos update} by
\begin{equation}
|\mathfrak{A}_{k+1}\rangle=\mathfrak{D}\cdot|r\rangle,\label{eq:David Referenz}
\end{equation}
where $|r\rangle\bot|\mathfrak{A}_{1}\rangle=|A_{[n]}^{[\textrm{refer}]}\rangle\approx|e_{0}^{[n]}\rangle\approx|e_{0}^{[n-1]}\rangle$
and with that $|e_{0}\rangle\langle e_{0}|r\rangle\approx0\cdot|e_{0}\rangle$.
Afterwards, residual parts of the $|e_{0}\rangle\langle e_{0}|$ term
are exfiltrated again because $|\mathfrak{A}_{k+1}\rangle$ has to
be orthogonalized (and normalized)
\begin{equation}
|\mathfrak{A}_{k+1}\rangle\longleftarrow\left(\mathbb{I}-\sum_{i=1}^{k}|\mathfrak{A}_{i}\rangle\langle\mathfrak{A}_{i}|\right)\cdot|\mathfrak{A}_{k+1}\rangle.
\end{equation}
These considerations are also part of the more elaborated Jacobi-Davidson
\cite{DavidsonJacobi} method to which this implementation can be
extended.

We might further consider to omit terms with $e_{i}\gg e_{0}$ since
their influence shrinks with $\left(e_{0}-e_{i}\right)^{-1}$. At
the end of the day, the effort to construct $\mathfrak{D}$ as well
as the effort for each application scale with $N$ times the number
of $|e_{i}\rangle\langle e_{i}|$ terms used in $\mathfrak{D}$.

\section{Altered minimization\label{sub:Appendix:-Altered-minimization}}

Here we derive the missing equations of \ref{sub:Translational-invariant-ground-1}.
First, we search for an approximation of $\lambda_{[L/R]}$ and start
by an alternative way of expressing them
\begin{equation}
\lambda_{[L]}^{\alpha\beta}=Q_{[L]\, s}^{*\,\alpha\gamma}\cdot A_{s}^{\gamma\beta};\quad\lambda_{[R]}^{\alpha\beta}=A_{s}^{\alpha\gamma}\cdot Q_{[R]\, s}^{*\,\gamma\beta},\label{eq:Lambda exakt}
\end{equation}
where we used the orthogonality \eqref{eq:OrthogonalQ} of the $Q$
and the decomposition \eqref{eq:Decomposition}. Next, we use the
fact that the algorithm is tuned to produce consecutive tensors $A_{[n-1]},A_{[n]}$
which are quite similar. Hence we approximate the yet unknown $Q_{[L/R]}^{*}$
\eqref{eq:Lambda exakt} by their known predecessor of the optimization
round before, i.e.\  $Q_{[L/R]}^{[n]\,*}\approx Q_{[L/R]}^{[n-1]\,*}$.
\begin{eqnarray}
\bar{\lambda}_{[L]}^{[n]\,\alpha\beta} & = & Q_{[L]\, s}^{[n-1]*\,\alpha\gamma}\cdot A_{s}^{[n]\,\gamma\beta}\nonumber \\
\bar{\lambda}_{[R]}^{[n]\,\alpha\beta} & = & A_{s}^{[n]\,\alpha\gamma}\cdot Q_{[R]\, s}^{[n-1]*\,\gamma\beta}.\label{eq:Approx Lambda}
\end{eqnarray}

Next, we use the same idea and replace the $Q_{[L/R]}^{[n-1]}$ in
definition \eqref{eq:Neue Trans Inv Basis Vektoren} by the $Q_{[L/R]}^{[n]}$
\begin{eqnarray}
|\mathfrak{\bar{A}}_{i}\rangle & \approx & \frac{1}{2}|\mathfrak{A}_{i}\rangle+\frac{1}{4}\left(Q_{[L]}^{[n]}\cdot\bar{\lambda}_{[R]\, i}+\bar{\lambda}_{[L]\, i}\cdot Q_{[R]}^{[n]}\right)\nonumber \\
 & \textrm{with} & \bar{\lambda}_{[R]\, i}^{\alpha\beta}\approx|\mathfrak{A}_{i}\rangle{}_{s}^{\alpha\gamma}\cdot Q_{[R]\, s}^{[n]*\,\gamma\beta}\nonumber \\
 &  & \bar{\lambda}_{[L]\, i}^{[n]\,\alpha\beta}\approx Q_{[L]\, s}^{[n]*\,\alpha\gamma}\cdot|\mathfrak{A}_{i}\rangle{}_{s}^{\gamma\beta},.
\end{eqnarray}
With that we like to calculate $|\bar{A}\rangle=|\mathfrak{\bar{A}}_{i}\rangle\cdot a_{i}$
\eqref{eq:Symmetrisch A}. Therefor we first observe
\begin{eqnarray}
\bar{\lambda}_{[L]\, i}^{[n]}\cdot a_{i} & \approx & Q_{[L]\, s}^{[n]*\,\alpha\gamma}\cdot|\mathfrak{A}_{i}\rangle{}_{s}^{\gamma\beta}\cdot a_{i}\nonumber \\
 & = & Q_{[L]\, s}^{[n]*\,\alpha\gamma}\cdot A_{s}^{\gamma\beta}\nonumber \\
 & = & \lambda_{[L]}^{[n]\,\alpha\beta},\label{eq:Summen Lambda}
\end{eqnarray}
where we used $A=|\mathfrak{A}_{i}\rangle\cdot a_{i}$ \eqref{eq:Nochmal linear combi}
in line two and equation \eqref{eq:Lambda exakt} in line three. Likewise
we find $\bar{\lambda}_{[R]\, i}\cdot a_{i}\approx\lambda_{[R]}$
and
\begin{eqnarray}
|\mathfrak{\bar{A}}_{i}\rangle\cdot a_{i} & \approx & \left[\frac{1}{2}|\mathfrak{A}_{i}\rangle+\frac{1}{4}\left(Q_{[L]}\cdot\bar{\lambda}_{[R]\, i}+\bar{\lambda}_{[L]\, i}\cdot Q_{[R]}\right)\right]\cdot a_{i}\nonumber \\
 & \approx & \frac{1}{2}|A\rangle+\frac{1}{4}\left(Q_{[L]}\cdot\lambda_{[R]}+\lambda_{[L]}\cdot Q_{[R]}\right)\nonumber \\
 & = & \frac{1}{4}|A\rangle+\frac{1}{4}|A\rangle+\frac{1}{4}Q_{[L]}\cdot\lambda_{[R]}+\frac{1}{4}\lambda_{[L]}\cdot Q_{[R]}\nonumber \\
 & = & \frac{1}{4}Q_{[L]}\cdot\lambda_{[L]}+\frac{1}{4}Q_{[R]}\cdot\lambda_{[R]}+\nonumber \\
 &  & +\frac{1}{4}Q_{[L]}\cdot\lambda_{[R]}+\frac{1}{4}\lambda_{[L]}\cdot Q_{[R]}\nonumber \\
 & = & \frac{1}{2}\left(Q_{[L]}\cdot\lambda_{[\textrm{sym}]}+\lambda_{[\textrm{sym}]}\cdot Q_{[R]}\right)\quad\textrm{with}\nonumber \\
\lambda_{[\textrm{sym}]} & = & \frac{1}{2}\left(\lambda_{[L]}+\lambda_{[R]}\right),
\end{eqnarray}
as claimed in equation \eqref{eq:Symmetrisch A}.

\section{Mirror symmetry\label{sub:Appendix:-Mirror-symmetry}}

Here we show that in case of a mirror symmetric\emph{ }Hamiltonian
$\mathcal{H}$ i.e.\  a Hamiltonian that is invariant under inversion
of the order of its sites\emph{ }
\begin{equation}
\mathcal{H}_{s_{1}\cdots s_{n}}^{s'_{1}\cdots s'_{n}}=\mathcal{H}_{s_{n}\cdots s_{1}}^{s'_{n}\cdots s'_{1}}\label{eq:InvertHam}
\end{equation}
all tensors $A_{[n]\, s}^{\alpha_{l}\alpha_{r}}$ can be chosen mirror
symmetrical in their auxiliary indices $\alpha_{l},\alpha_{r}$, i.e.
\begin{equation}
A_{[n]\, s}^{\alpha_{l}\alpha_{r}}=A_{[n]\, s}^{\alpha_{r}\alpha_{l}},\label{eq:SpeigelSymA}
\end{equation}
This allows to impose an extra constraint on $A_{[n]}$. Commuting
the indices $\alpha_{l},\alpha_{r}$ also results in
\begin{eqnarray}
Q_{[L]\, s}^{\alpha_{l}\alpha_{r}} & = & Q_{[R]\, s}^{\alpha_{r}\alpha_{l}}\nonumber \\
\lambda_{[L]}^{\alpha_{l}\alpha_{r}} & = & \lambda_{[R]}^{\alpha_{r}\alpha_{l}},\label{eq:Spiegel Q}
\end{eqnarray}
as can be seen directly from the decomposition \eqref{eq:Decomposition}.
Further it is possible to construct $L^{\alpha'_{l}\mu_{l}\alpha_{l}}$
and $R^{\alpha'_{r}\mu_{r}\alpha_{r}}$ \eqref{eq:New Lef Right Half}
such that they are identical. But therefor we need to resort to an
alternative MPO construction for the Hamiltonian, such that the MPO
tensors of the left half are mirror symmetric to the tensors of the
right half. This can be achieved if we include a special interface
tensor in the middle where $L^{\alpha'_{l}\mu_{l}\alpha_{l}}$ and
$R^{\alpha'_{r}\mu_{r}\alpha_{r}}$ are connected to build $\mathbb{\widetilde{H}}$.
This reduces the requirement in storage memory roughly by a factor of 2, but has nearly no effect on the speed. Since storage capacities
are usually not a big issue, we do not elaborate this point any further.

One should be aware that equation \eqref{eq:InvertHam} enforces a mirror symmetric MPS.
In case of a mirror symmetry breaking ground state, the MPS will represent a superposition of both chiralities,
implying an unfavorably increased requirement in bond dimension.

We further remark that our definition of a mirror symmetric Hamiltonian
does not forcedly imply a symmetry in real space. Although in practical
application the order of the sites generally coincides with one specific
spatial direction, there is no mathematical connection between the
direction of space and the chosen order.

Assuming a mirror symmetric Hamiltonian $\mathcal{H}$ \eqref{eq:InvertHam}
the claim of the mirror symmetric tensor $A_{[n]\, s}^{\alpha_{l}\alpha_{r}}$
\eqref{eq:SpeigelSymA} can be proven iteratively:
\begin{enumerate}
\item If $A_{[i]\, s}^{\alpha_{l}\alpha_{r}}=A_{[i]\, s}^{\alpha_{r}\alpha_{l}}$
for all $i<n$, then $\mathbb{\widetilde{H}}_{[n]}$ \eqref{eq:Sum Min}
is mirror symmetric, i.e.\  $\mathbb{\widetilde{H}}_{[n]\, ss'}^{\alpha'_{l}\alpha_{l}\alpha'_{r}\alpha{}_{r}}=\mathbb{\widetilde{H}}_{[n]\, ss'}^{\alpha'_{r}\alpha_{r}\alpha'_{l}\alpha{}_{l}}$.
\item If 1. is fulfilled, then $A_{[n]\, s}^{\alpha_{l}\alpha_{r}}=A_{[n]\, s}^{\alpha_{r}\alpha_{l}}$
\eqref{eq:SpeigelSymA}.
\end{enumerate}
The very first $\mathbb{\widetilde{H}}_{[1]}$ is constructed via
the initialization procedure described in \ref{sub:Initialization}.
If we start with a mirror symmetric wave function and use Takagi's
factorization as suggested in 3. of the initialization procedure,
$\mathbb{\widetilde{H}}_{[1]}$ is symmetric and the induction is
well grounded.

\emph{Proof for 1}. All $\mathbb{\widetilde{H}}$ represent a sum
of operators according to equation \eqref{eq:Sum Min}
\begin{equation}
\mathbb{\widetilde{H}}=\sum_{i=1}^{4^{n}}\mathbb{H}_{i}.
\end{equation}
Each of these $\mathbb{H}_{i}$ contains an entire Hamiltonian MPO as
central unit sandwiched by bra and ket MPS, with a hole where the
new tensor $A_{[\textrm{new}]}=A_{[n]}$ is supposed to be inserted.
The Hamiltonian is guaranteed to be mirror symmetric, while there
is no such condition for the MPS. The different MPS are encoded in
the building blocks $L^{\alpha'_{l}\mu_{l}\alpha_{l}}$ and $R^{\alpha'_{r}\mu_{r}\alpha_{r}}$,
which are constructed symmetrically \eqref{eq:New Lef Right Half},
i.e.,\  in contrast to the basic algorithm each new tensor is inserted
in $L^{\alpha'_{l}\mu_{l}\alpha_{l}}$ and $R^{\alpha'_{r}\mu_{r}\alpha_{r}}$
at equal footing. Hence, for each MPS exists a counterpart which contains
exactly the same tensors in inverted order. In general, this alone
is not enough, because mirror symmetry also exchanges the left and
right auxiliary indices. But, since all involved tensors $A_{[i]\, s}^{\alpha_{l}\alpha_{r}}=A_{[i]\, s}^{\alpha_{r}\alpha_{l}}$
are supposed to be invariant under this kind of exchange (and $Q_{[L]\, s}^{\alpha_{l}\alpha_{r}}=Q_{[R]\, s}^{\alpha_{r}\alpha_{l}}$
\eqref{eq:Spiegel Q}, as needed), mirror symmetric counterparts for
all involved MPS are guaranteed and with that $\mathbb{\widetilde{H}}$
is mirror symmetric.

\emph{Proof for 2}. The tensor $A_{[n]\, s}^{\alpha_{l}\alpha_{r}}$
is the result of the minimization procedure described in more details
in \ref{sub:Minimization-routine-and}. Each element in this procedure
maintains mirror symmetry if $\mathbb{\widetilde{H}}_{[n]}$ and the
initial tensor $A_{[n]}^{[\textrm{refer}]}$ \eqref{eq:InitA} are
mirror symmetric. Since $A_{[n]}^{[\textrm{refer}]}$ is a superposition
of mirror symmetric tensors, all conditions are met.

Finally we remark that accumulating numerical errors might undermine
the symmetry. Therefore we recommend to explicitly restore the symmetry
of each $A_{[n]\, s}^{\alpha_{l}\alpha_{r}}$ during its calculation.

\bibliographystyle{apsrev4-1}
\bibliography{Zitate}

\end{document}